\documentclass[conference]{IEEEtran}

\makeatletter
\def\ps@headings{%
\def\@oddhead{\mbox{}\scriptsize\rightmark \hfil \thepage}%
\def\@evenhead{\scriptsize\thepage \hfil \leftmark\mbox{}}%
\def\@oddfoot{}%
\def\@evenfoot{}}
\makeatother
\usepackage{times,cite}
\usepackage{graphics}
\usepackage{graphicx}
\usepackage{subfigure}
\usepackage{amssymb}
\usepackage{amsmath}
\usepackage{amsfonts}         
\usepackage{color}            
\usepackage{epic}             
\usepackage{eepic}            
\usepackage{rotating}         
\usepackage{type1cm}          
\usepackage{epsfig}
\usepackage{amsthm}
\usepackage{amsmath}
\usepackage{graphicx}
\usepackage{graphics}
\usepackage{epsfig}
\usepackage{latexsym}
\usepackage{amsfonts}
\usepackage{amssymb}
\usepackage{paralist}
\usepackage{xspace}
\usepackage{mathrsfs}
\usepackage{psfrag}
\usepackage{color}
\usepackage[ruled]{algorithm}
\usepackage[noend]{algorithmic}

\input epsf

\newcommand{\ie}{\textit{i.e.}\xspace}
\newcommand{\eg}{\textit{e.g.}\xspace}

\def\ie{\textit{i.e.}\xspace}

\def\etc{\textit{etc.}\xspace}

\def\eg{\textit{e.g.}\xspace}

\newcommand{\equref}[1]{{Eq.~(\ref{#1})}}

\def\ourprotocol{\textbf{SilentSense\ }}
\def\ourprotocoltight{\textbf{SilentSense}}

\newcommand{\p}{\mathbf{P}}
\newcommand{\delay}{\mathbf{d}}

\begin{document}
\title{SilentSense:  Silent User Identification via  Dynamics
  of  Touch and Movement  Behavioral Biometrics}

\author{\IEEEauthorblockN{Cheng Bo\authorrefmark{1}, Lan Zhang\authorrefmark{2}, Xiang-Yang Li\authorrefmark{1}}
\IEEEauthorblockA{\authorrefmark{1}Department of Computer Science, Illinois Institute of Technology, USA\\
\authorrefmark{2}Department of Software Engineering, Tsinghua University, China PR}
Email: cbo@hawk.iit.edu, zhanglan03@gmail.com, xli@cs.iit.edu\vspace{-0.2in}
}

\maketitle

\begin{abstract}
With the increased popularity of smartphones, various security threats
 and privacy leakages targeting them are discovered and investigated.
In this work, we present \ourprotocoltight, a framework to authenticate
 users silently and transparently by exploiting dynamics mined from
 the  user touch  behavior biometrics and
 the micro-movement of the device caused by user's
 screen-touch actions.
We build a ``touch-based biometrics'' model of the owner by
 extracting some principle features, and then verify whether the
 current user is the owner or  guest/attacker.
When using the smartphone,
 the unique operating dynamics of the user is detected and learnt by
 collecting the sensor data and touch events silently.
When users are mobile, the micro-movement of mobile devices caused by
 touch is suppressed by that due to the large scale user-movement
 which will render the  touch-based biometrics ineffective.
To address this, we integrate a movement-based biometrics for each
 user with previous touch-based biometrics.
We conduct extensive evaluations of our approaches on the Android
 smartphone, we show that the user identification accuracy is
 over $99\%$.
\end{abstract}

\begin{keywords}
Identification, Touch, Behavioral Biometrics,  Security.
\end{keywords}

\section{Introduction}
\label{sec:intro}
The rapid development in mobile device industry has now stimulated the
 blooming of the personalized applications (checking email, enjoying personal
 photos) and services (mobile payment, smart home~\cite{vu2012distinguishing})
 for more convenience and better user experience.
As mobile devices get involved in more and more parts of people' daily lives,
 the device owner faces increasing risk of privacy leakage not only from the system or apps,
 but also from sharing device with guest users, such as family members,
 friends, partners, or coworkers.
According to survey~\cite{karlson2009can}, a large number of users concern about
 their data privacy and integrity when sharing mobile phones to others.
User identification works as an important component
 of mobile devices for personalized services and data access control.
However, explicit identification mechanisms, \eg PIN code or password,
are annoying, and not application/content specific.
Besides, before handing the device to the guest user,
 a owner-triggered protection strategy is labor-intensive
 and makes most owner awkward because it shows distrust to the guest \cite{karlson2009can}.
Under this circumstance, it would be good for device to identify
 the current user swiftly, silently, and inconspicuously, as
 well as provide necessary privacy protection and access control automatically.

At present, PIN codes or password is the most common identification and access control
 strategy in commercial mobile device operating systems, such as iOS,
 which is obviously labor-intensive and time-consuming.
Facial recognition~\cite{applock} by the front camera is another optional strategy
 to identify users.
But it is still annoying to require users to take pictures frequently.
Besides, the face recognition accuracy is unreliable with changing
environment and
 frequent imaging is power-consuming.
The latest solution exploits the capacitive touch communication as a
 mechanism to distinguish different users~\cite{vu2012distinguishing},
 which utilizes touch screen device as receivers for an identification
 code transmitted by a hardware identification token.
This mechanism requires special devices.
All these mechanisms have the risk of being imitated,
 \eg, peeking at the PIN code, using a photo to cheat the camera, or
 eavesdropping on the communication between the transceivers.

On the other hand, the motion sensors integrated in most modern smartphones have
 stimulated the research on user behavior detection.
For example, TapPrints~\cite{miluzzo2012tapprints} indicates the behavior of tapping on different
 locations on the touch screen will be reflected from sensory data, and such
 observation may be considered as potential risk of compromising user's privacy~\cite{cai2009defending}.
In addition, individual users may have their own interacting behavior patterns,
 and these motion sensors may help to characterize user's behavior to identify users~\cite{WM-CS-2012-06}.

In this work, investigating the feasibility of utilizing the behavioral
 biometrics extracted from smartphone sensors for user identification,
 we propose \ourprotocoltight, a non-intrusive
 user identification mechanism to silently substantiate whether the current
 user is the device owner or a guest or even an attacker.
The interacting behavior could be observed while the smartphone is in
 relatively static condition.
Exploiting the combination of several interacting features from both touching behavior
 (pressure, area, duration, position) and reaction of devices (acceleration and rotation),
 \ourprotocol  achieves highly accurate identification with low delay.
A great challenge comes from the circumstance when the user is in motion, such as walking.
 The perturbation generated by the interacting will be suppressed
 by larger-scale user movement.
While most of existing works neglect this circumstance,
 \ourprotocol is capable of identifying user in motion by extracting
 the motion behavior biometrics.
As long as the current user is identified, necessary access control is
 triggered automatically.

Keeping the sensors always on
 provides minimum guest identification delay,
 but could cause unwanted energy consumption
 since most of time the current user is the owner or the current app is not sensitive, \eg, the game app.
But an intrusion may be missed when the sensors are off.
Facing this debacle,
 we propose a novel model
 to estimate the current user leveraging the observation of owner's sociable habit.
An online decision mechanism
 is designed for the timing to turn on or turn off sensors,
 which provides a balance between energy cost, delay and accuracy.
Our online decision
 mechanism results in an adaptive observing frequency
 according to owner's social habit.

We evaluate the effectiveness of \ourprotocol through
 extensive experiments using the empirical data collected from
 100 volunteer users in both static and motion scenarios.
The evaluation results indicate that in the former scenario,
 our approach is able to classify the
 legitimate user and guest/impostors with averaged equal error rate
 about $20\%$ with only one stroke, $0$ error with about $13$ strokes.
However, when users are moving, the approach designed for static
 scenario deteriorate to with false reject ratio (FRR)
 about only $18 \%$ after 4 steps.
Our integrated approach using motion behavior biometrics improves the
 FAR and FRR to reaches $0$ with only $3$ steps.
Our study also shows that individual  behavior patterns are difficult
 to be imitated precisely, and the feature is unique.
The evaluation also shows our online decision mechanism
 can successfully identify the guest user with averaged 2.26 actions delay
 with a $98\%$ accuracy guarantee when $57\%$ time the sensors are off.

The rest of the paper is organized as follows.
In Section \ref{sec:overview} we provide the overview of our
 approaches.
We discuss in detail how to detect various user behaviors in
Section~\ref{sec:detection} and
 present the performance evaluation of our approaches in
Section~\ref{sec:evaluation}.
We review the related work in Section~\ref{sec:related},
and conclude the paper in Section~\ref{sec:conclusion}.

\section{System Overview}
\label{sec:overview}
\ourprotocol is designed as a pure software-based framework, running in
 the background of smartphone, which non-intrusively explores the behavior
 of users interacting with the device without any additional assistant hardware.

\subsection{Main Idea}
The main idea of \ourprotocol for user identification comes from two aspects:
(1) how you use the device; and (2) how the device reacts to the user action.

While using mobile devices, most people may follow certain individual habits
 unconsciously.
Running as a background service,
 \ourprotocol exploits the user's app usage and interacting behavior with each app,
 and uses the motion sensors to measure the device's reaction.
Correlating the user action and its corresponding device reaction,
 \ourprotocol establishes a unique biometric model to identify the
 role of current user.

We investigate the phone usage behavior of our colleagues.
Tiny perturbation of the whole device will be captured by motion sensors
 when a user touches the screen.
The amplitude of such tiny perturbation depends on the user's
 holding gesture, the touching pressure and coordinate.
The framework focuses on extracting essence features
  from the user's behavior, including both screen-touch events and the
  user's motion events,
  to determine the discriminative patterns of individuals.
Such behavior pattern and dynamics are much difficult to be imitated
or attacked as these are often invisible.
Besides, both our investigation and \cite{karlson2009can} show
 that people share phone with friends from time to time
 and the share frequency varies with the owners' social habit.
The phone belonging to a more sociable owner,
 tends to have a higher probability to be shared with guest users,
 which may require a relatively high identification frequency, and vice versa.
The social characteristic of the owner can help us to
 optimize the observation frequency to reduce the overall energy cost
 with a identification performance guarantee.

While the owner is using the phone,
 it is feasible to establish a behavior model through
 automatically learning.
When interacting happens, the system evaluates
 the probability of being the owner,
 and updates the evaluation
 with increasing observations to determine the identity silently and
 automatically.
If the current user is a guest, the privacy protection
 mechanism will be triggered automatically, which
 prevents the privacy leakage while maintaining the
 trustiness of the guest user.
Based on the historical identification results,
 the social characteristic of the owner could be learned
 to help decide the observation frequency.

\subsection{Challenges}
In order to achieve the identification in an accurate, silent and fast manner,
 the following technical challenges should be addressed.
\begin{compactdesc}
\item[User Behavior Modeling:]
To characterize individuals' unconscious use habits accurately,
 the user behavior model should contain multiple features
 of both user's action and device's reaction.
In addition, the connection between features may not be neglected for identification,
 such as different interaction coordinates on the touch screen
 may cause different reaction vibrations of the device.
\item[Identification Strategy:]
To establish the user behavior model,
 for the owner there are abundant behavior information.
For a guest, the collected behavior information may be very limited.
In addition, in motion scenarios, some interacting features will be
 swamped by the motion from the perspective of sensory data,
 which greatly increases the difficulty of accurate identification.
Thus it is challenging to distinguish users with limited information effectively,
 even if the interacting features are partially swamped.

\item[Balance Among Accuracy, Delay and Energy:]
Nonstop observation with sensors provides identification with small delay
 and high accuracy,
 but may cause unwanted energy consumption for the mobile device when the current user
 is the owner or a guest is using an insensitive app, \eg, playing a game.
But intrusion may happen when the sensors are off and the risk increases with the detection delay.
A well designed mechanism is required to decide the observation timing
 to reduce energy consumption while guarantee the identification accuracy and delay.


\end{compactdesc}

\subsection{Main Framework}
\begin{figure}[!hptb]
\centering
\includegraphics[width=3.5in]{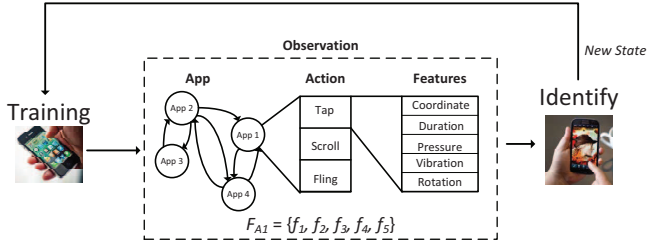}
\caption{Framework Overview.}
\label{fig:overview}
\end{figure}
The framework model consists of two basic phases: \emph{Training}
 and \emph{Identification}, as shown in Figure~\ref{fig:overview}.
The training phase is conducted to build a behavior model when the user
 is interacting with the device, and the identification phase is implemented
 to distinguish the identity of the current user based on the observations
 of each individual's interacting behaviors.
When a guest user is observed,
 privacy protection mechanism will be triggered automatically.
After the guest leaves and the owner returns,
 privacy protection will be reset for the owner's convenience.

Initially, we assume that the device has only the owner's information,
 \eg a newly bought phone.
The framework trains the owner's behavior model by retrieving two
 types of correlated information, the information of each touch-screen action
 and the corresponding reaction of the device when the user is static.
In the motion scenario,
 instead of the reaction information,
 motion features will be detected.
With the owner's behavior model,
 the current user will be identified through one-class SVM classification.
One-class SVM provides a judgement whether the observed features belong
to the owner (true of false).
However the identification accuracy by one observation
 is usually not high enough for an identity conclusion,
 continuous consistent judgements will increase the accumulated
 confidence for this judgement.
When the accumulated confidence is high enough,
 a conclusion is ready and the newly observed features will be added to the owner or
 guest dataset according to the conclusion to update the model.
Gradually, by self-learning,
 the model will be upgraded to two-class SVM model,
 which provides more accurate judgement.

To reduce energy cost of frequent observations,
 an optimal stoping mechanism is designed to determine
 the timing to stop observation \ie, turn sensors off, with a good
 accuracy guarantee.
Based on the recent historical conclusions,
  the social characteristic of the owner can be learnt.
Here the social characteristic refers in particular to
 the frequency/propobility this owner shares
 his/her device with a guest.
With the help of the social characteristic,
 a strategy is designed to determines the timing to restart observation.

\subsection{Interacting Model}
For touch actions,
 there are three principle gestures: \emph{tap}, \eg, texting, clicking item,
 \emph{scroll}, \eg, browsing mails and tweets, and \emph{fling}, \eg, reading  e-books.
Different gestures usually have different touch features
 and lead to different device reactions.
Interacting with certain app often involves a certain set of gestures.
For a touch action $T_i$,
 we combine the app with its touch gesture and the
 features captured by this framework as one \emph{observation},
 denoted as $O_i = \{A_i, G_i, f_{i1}, \cdots, f_{in}\}$.
 here $A_i$ is the app being used, $G_i$ represents the
 gesture (\eg tapp), and $f_{i,j}$ ($j \ge 1$) are features of the observed action.

Because of individual habits,
 the features of the same gesture for the same app vary for different users.
Two types of features are used in this system:
 the \emph{touch} features and \emph{reaction} features.
The touch features include touch coordinate on the screen, touch pressure and duration,
 which can be obtained from system API.
To capture the reaction features,
 we notice that diverse gestures and positions for holding the device by individual
 users infer different amplitudes of vibration caused by each touch,
 which has already been proved by previous works (\cite{xu2012taplogger,
  miluzzo2012tapprints}).
Such tiny reaction of the devices produces an identifiable patterns
 which could be observed via accelerometer and gyroscope.
Therefore, for each gesture ($G_i$) in one observation,
 its feature is the combination of three touch features and two reaction features,
 presented as: $F_{G_i} = \{f_1, f_2, f_3, f_4, f_5\}$.

\subsection{Identification Strategy}
Both the training and identification process are established based on
 observations from interacting behavior as illustrated in
 Figure~\ref{fig:overview}.
In this section,
 we present our self-learning model and identification strategy.

Initially, without only the owner's behavior data,
 a one-class SVM model is trained to identify a new observation $O_i$
 and provide the judgement $J_i$ whether this action belongs to
 the owner or not, \ie, \emph{$J_i=true$} or \emph{$J_i=false$}.
Lacking of groundtruth,
 it is difficult to determine the correctness of the judgement.
To achieve high identification accuracy,
 we adopt the SVM model's credibility for each judgment $J_i$ as the \emph{confidence}
 of the framework on $J_i$.
Let this confidence be $\varepsilon_i(J_i)$,
 which indicates the probability the framework considers that $J_i$ is
 correct for the current observation $O_i$.
Using one-class SVM,
 the judgement of one observation usually is
 not accurate enough to make a identity conclusion.
Obviously, more observations leads to higher conclusion accuracy.
Specifically,
 let $\{J_1, J_2,\cdots,J_k\}$ be a sequence of consistent judgements, \ie $J_1=J_2=\cdots=J_i$,
 to continuous observations $\{O_1, O_2,\cdots,O_k\}$.
Based on the judgement sequence,
 an identity conclusion $I_{1,k}$ can be made, \ie $I_{1,k}=J_k$.
Then the \emph{conclusion confidence} will be cumulated as
 \begin{equation}\label{eq:cc}
 \p_{1,k} = \p(J_1, J_2,\cdots,J_k) =1 - (\prod\limits_{i = 1}^{k} (1 -
 \varepsilon_i(J_i))),
 \end{equation}
 which indicates the probability this framework considers that the identity
 conclusion $I_{1,k}$ is correct.
Then the identification delay $\delay_k$ for a conclusion $I_{1,k}$
 is defined as the number of observations taken to achieve this conclusion.
With the number of observation increases, the framework will be more
 confident to provide a correct conclusion, meanwhile the delay will increase.
Note that,
 an inconsistent judgement will interrupt the sequence,
 and the conclusion confidence need to be cumulated from scratch.
Except some multi-player game, which is not privacy sensitive,
 in most cases,
 there won't be frequent switches between guest and owner.
Since a conclusion will change the privacy setting,
 a high confidence is required to give an identity conclusion.
A conclusion confidence threshold $\p_\theta$ can be given to make sure
 this framework only outputs conclusion with confidence higher than $\p_\theta$.

While the system uses the owner's model to make identification conclusions,
 the observation generating a judgement with high confidence
 will be added into the owner or guest training data buffer accordingly.
With a small amount of guest data,
 our system upgrades the model to \emph{two-class} SVM model,
 which outperforms the one-class model in accuracy.
The SVM model will be continuously updated using the most recently
buffered data.
Considering the training data, the number of owners usually will be
 far greater than the number of guest users,
 the updating frequency of guest behavior should be higher than that
 of owner's.

\subsection{Observation Decision}
In practice,
 user usually requires a high conclusion confidence,
 which may lead a undesired long conclusion delay.
In addition,
 the main energy cost of \ourprotocol is caused by sensors.
Nonstop observation may cause unwanted energy consumption,
 as for most of the time the user is the owner.
The balance among accuracy, delay and energy is challenging.
In this section, we investigate the requirements of the device
 owner and design a strategy to determine the timing to start and
 stop observation,\ie turn on and turn off sensors.

First, we consider the privacy requirements of the owner.
Not all apps are privacy related,
 \eg, game apps are frequently used but nonsensitive.
For an sensitive app,
 if the user is a guest,
 the privacy protection mechanism should be enabled immediately;
 if the user is the owner,
 the protection mechanism should be disabled for the owner's convenience.
So if the current app is sensitive,
 the privacy and functionality requirements take priority over energy.
The current app can be detected by system API without extra energy cost.
Then we get the first rule of the strategy:
\emph{When there is a switch from a nonsensitive app to a sensitive app,
  observation should be started immediately (if hasn't bee).
And while the user is interacting with a sensitive app,
  the observation should not be stopped.}

When the current app is nonsensitive,
 observations help the framework to collect training data
 and get ready for a sudden app/user switch.
In this case, nonstop observation is not necessary and energy waste.
The issue is when to start/stop observation.
To guarantee a highly accurate conclusion with small delay,
 the main idea of our decision strategy is to keep the framework confident
 enough about the current user's identity using the sensors as less as possible.

When the observation with sensors has been started
 but a conclusion hasn't been made,
 the confidence of the current user's identity is the accumulated confidence
 according to \equref{eq:cc}.
Although larger number of observation leads to higher confidence,
 which implies higher accuracy,
 the energy cost will increase too.
Formally,
 let the current touch action be $T_k$, which can be acquired from system API,
 requiring no extra energy.
Its corresponding observation
 is $O_k$ and the judgement by SVM is $J_k$.
 Let $O_s$ be the observation from which the judgements are consistent with
 $J_k$, \ie, $J_i=J_k$ for $s<i<k$.
 Then the framework has a candidate identity conclusion $I_{s,k}=J_k$,
 and its confidence is $\p_{s,k} = 1 - (\prod\limits_{i = s}^{k} (1 -
 \varepsilon_i(J_i)))$.
The accumulated energy consumption from $O_s$ to $O_k$ is $E_{s,k}=\sum_{i=s}^ke_i$,
 here $e_i$ is the energy cost of sensors for observation $O_i$.
If the candidate identity conclusion is output after the action $T_k$,
 then the delay of the conclusion is $\delay_k = k-s$ (action).
We notice that there is a positive correlation between the delay and the energy cost,
 and both of them are preferred to be small.
Therefore, we propose $U_k=\p_{s,k}/E_{s,k}$ as the utility at the action $T_k$,
 which indicates a tradeoff among accuracy, energy and delay.
Then, deciding the time to stop observation can be solved as
 an Optimal Stopping problem, where after which touch action $T_t$
 to stop observation so that the utility is maximized and the conclusion confidence is guaranteed above $\p_\theta$.
We assume $R(U_t|\p_{s,t}>\p_\theta)$ as the expected maximum utility could be achieved by the following observation with a confidence constraint $\p_\theta$,
 which is:
\begin{eqnarray*}
\begin{split}
&R(U_t | \p_{s,t}>\p_\theta) = \\& \max \{{R(\frac{\p_{s,t}}{E_{s,t}}| \p_{s,t}>\p_\theta), R(\frac{\p_{s,t+1}}{E_{s,t+1}}| \p_{s,t+1}>\p_\theta)}\}.
\label{eq:dynamic_utility}
\end{split}
\end{eqnarray*}
Here, $P_{s,t+1}$ is the accumulated confidence by current confidence $P_{s,t}$ and the expected confidence $\bar\varepsilon$
 of SVM using \eqref{eq:cc}.
$E_{s,t+1} = E_{s,t} + \bar e$, here $\bar e$ is the expected energy cost for each observation.
Both $\bar\varepsilon$ and $\bar e$ can be obtained by the historical observations.
Once a stop timing $T_t$ is detected,
 the observation will be terminated,
 and a identification conclusion $I_{s,t}$ with confidence $\p_{s,t}>\p_\theta$ will be output for the privacy protection setting.

When the observation is stopped after the action $T_t$,
 the framework need to decide when to restart the observation.
As we mentioned,
 a switch to sensitive app will trigger the observation.
When the app is nonsensitive,
 we estimate the current user for action $T_j$ based on the recent conclusion $I_t$.
Intuitively, if $j-t$ is small enough, then with a high probability,
 the current user's identity is still consistent with $I_t$;
 as time goes by, the confidence of the identification $I_t$ will decrease.
In this work, we use the recent social state of the owner,
 which decides the confidence decrease rate,
 to estimate the current user.
When the confidence of estimation fall out of a lower bound $\p_\varphi$,
 the observation will be started.
Formally, the framework learns the transfer model between guest and
 owner from the recent historical conclusions.
The model includes two probabilities $Q_{o2g}$ and $Q_{g2o}$,
 which represent the transfer probability from \emph{owner} to \emph{guest}
 and \emph{guest} to \emph{owner} respectively.
When the observation is stopped,
 the identification for touch $T_j$ is consistent with the most recent conclusion $I_t$,
 with a confidence $\p_j = \p_{j-1}\cdot(1-Q_{o2g}) + (1 - \p_{j-1})\cdot{Q_{g2o}}$,
 if $I_t$ is owner; or $\p_j = \p_{j-1}\cdot(1-Q_{g2o}) + (1 - \p_{j-1})\cdot{Q_{o2g}}$,
 if $I_t$ is guest.
Once $\p_j<\p_\varphi$, the observation will be started,
 and a new round of optimal stopping is initiated.

Our online decision strategy achieves a tradeoff among accuracy, energy
 and delay with a confidence guarantee.
Meanwhile an adaptive observation frequency is yielded
 according to the owner's social habit.

\section{Behavioral Biometric Extraction}
\label{sec:detection}
Accurate behavioral biometric of the user
 is the core for correct identification.
Among the multi-dimension features,
 the pure touch features (coordinate, pressure and duration)
 are relatively easy to obtain via system API.
The challenge comes from how to
 capture the essence of the device reactions to
 each user's different actions,
 and the essence of a user's motion, \eg walking,
 using the noisy sensory values.
In this section, we will briefly introduce the
 our strategy to accurately extract the device reaction features
 and motion features.

\subsection{Device Reaction to Touch Action}
We use the onboard sensors
 to explore the device reaction to the three types of touch gestures.

We start the analysis from the features caused by tapping.
When tapping event happens,
 the tapping coordinate, timestamp and duration
 could be obtained from system API.
Meantime,
 the device accelerations along three device axes (X,Y,Z) are captured
 by the accelerometer.
In order to measure the amplitude of vibration while tapping,
 we use the $F_{tap} = \sqrt{{LA_x}^2 + {LA_y}^2 + {LA_z}^2}$
 to represent the summation of acceleration vector in the space.
$LA_x$, $LA_y$, $LA_z$ indicate the
 linear acceleration in the device system.
Another valuable reaction feature is the vector of angular velocity
 obtained from the gyroscope,
 denoted by $AV_{tap} = \sqrt{{AV_x}^2 + {AV_y}^2 + {AV_z}^2}$.
This feature represents the position variation of the
device in the space when touched by a user.

\begin{figure}[t]
\begin{center}
\subfigure[Accelerometer in Tap\label{fig:touch_acc}]{\includegraphics[scale = 0.25]{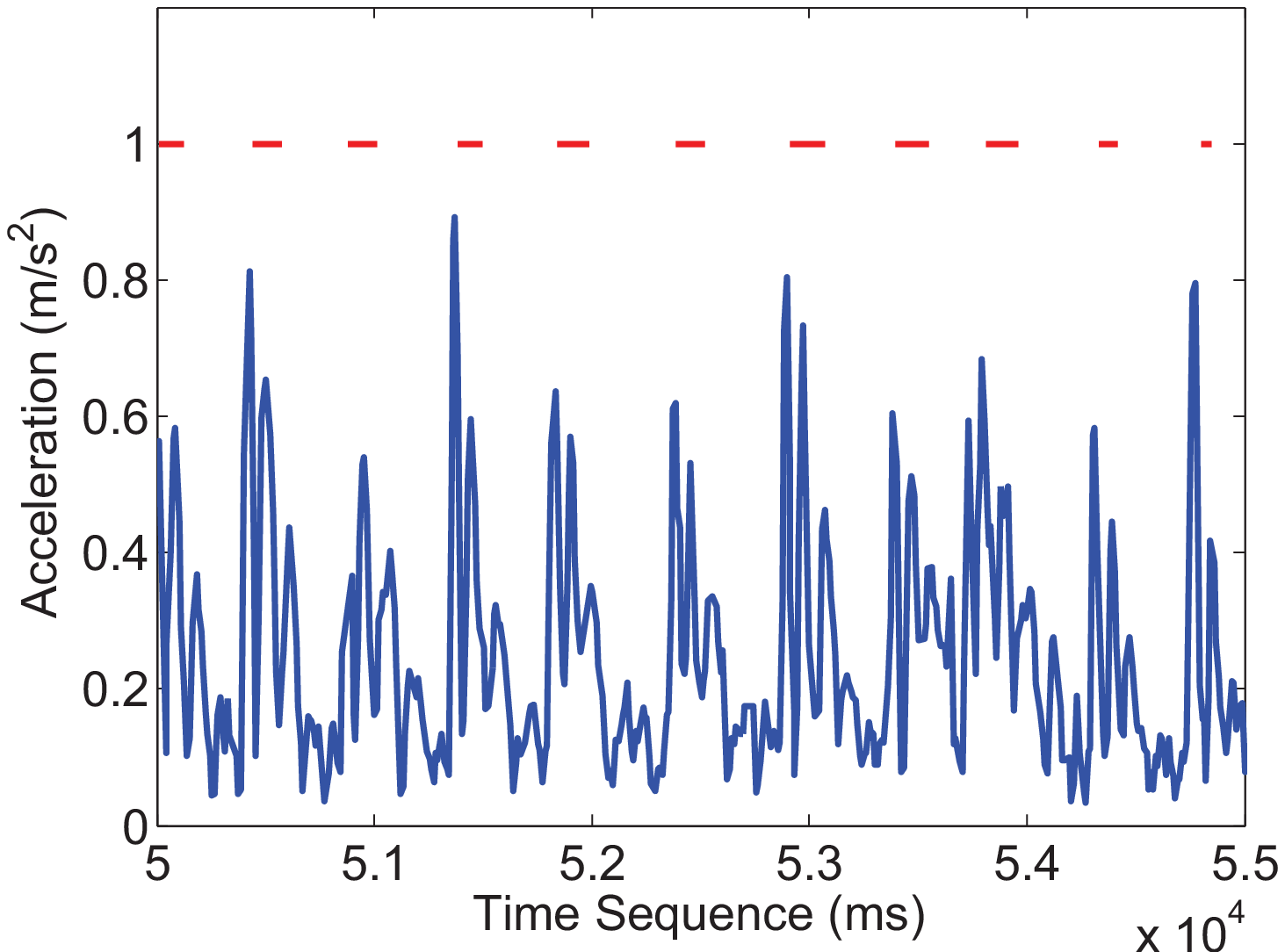}}
\subfigure[Gyroscope in
  Tap\label{fig:touch_gyro}]{\includegraphics[scale =
    0.25]{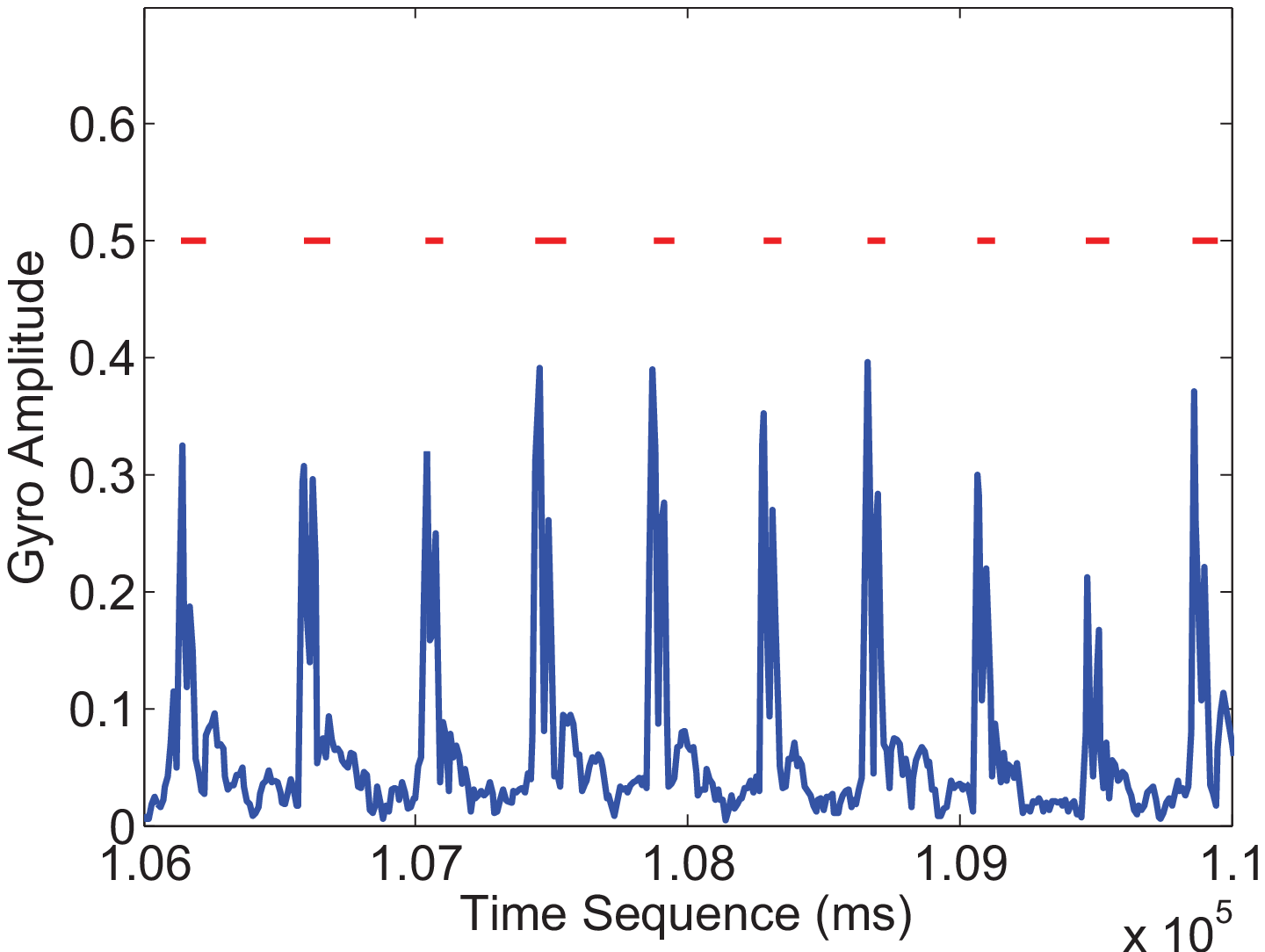}}
\caption{The reaction of the device when tapped.}
\label{fig:touch_acc_gyro}
\end{center}
\vspace{-0.15in}
\end{figure}

Figure~\ref{fig:touch_acc_gyro} illustrates the reaction of the mobile device
 when tapping event occurs while the user is sitting still.
In both sub-figures, the red line segments represents the occurrences of tapping,
 and the length of each segment corresponds to the duration of the tap event.
Tapping on the screen will cause jumping on the sensory data.
However, the sensory data contains noises and errors,
 because the user cannot hold the device in absolutely still.
Therefore,to eliminate such noise,
 we calculate the mean perturbation of acceleration and rotation as features
 for each tapping event.

\begin{figure}[t]
\begin{center}
\subfigure[Vibration in Tap\label{fig:dis_acc}]{\includegraphics[scale = 0.32]{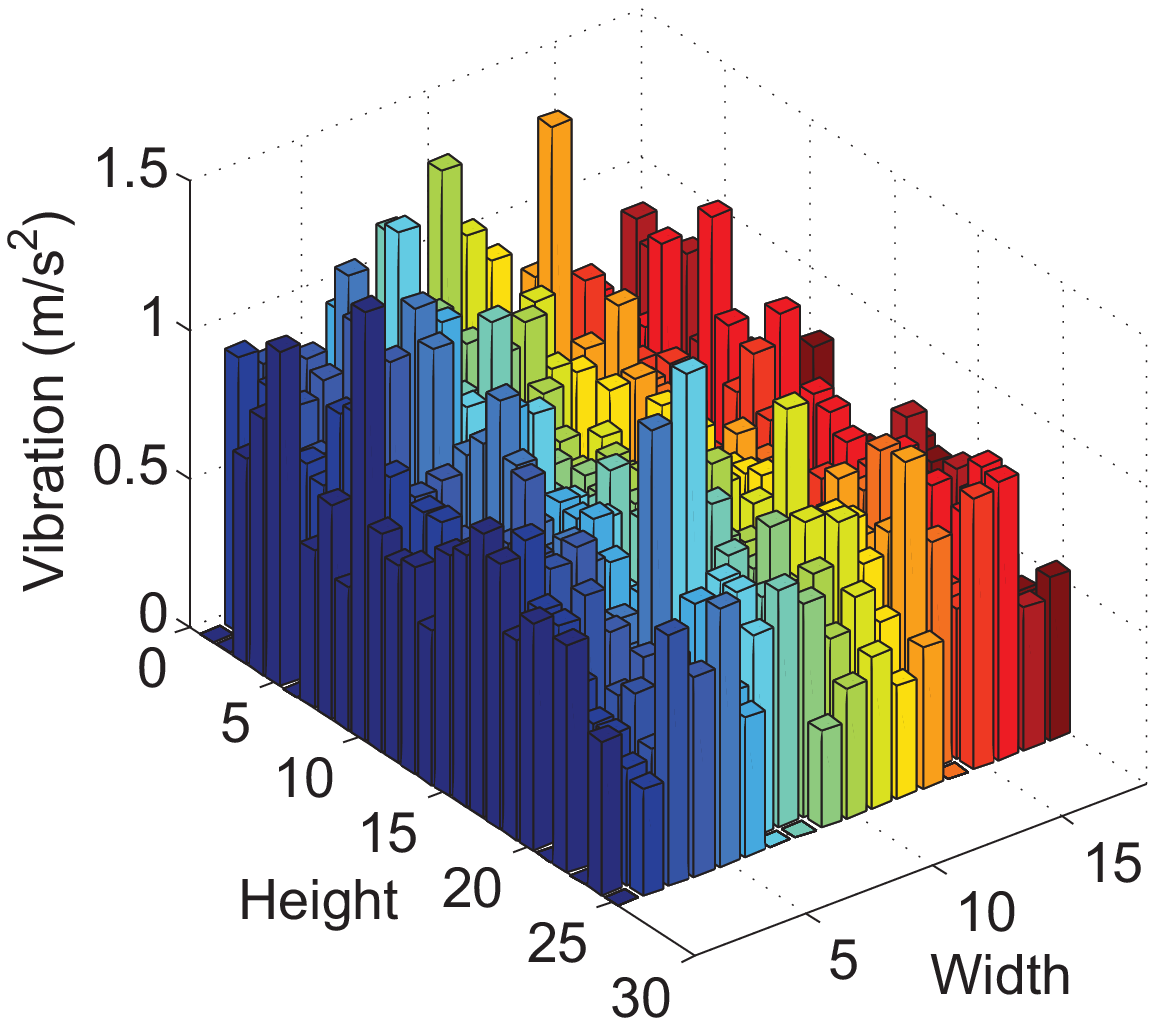}}
\subfigure[Rotation in Tap\label{fig:dis_gyro}]{\includegraphics[scale = 0.32]{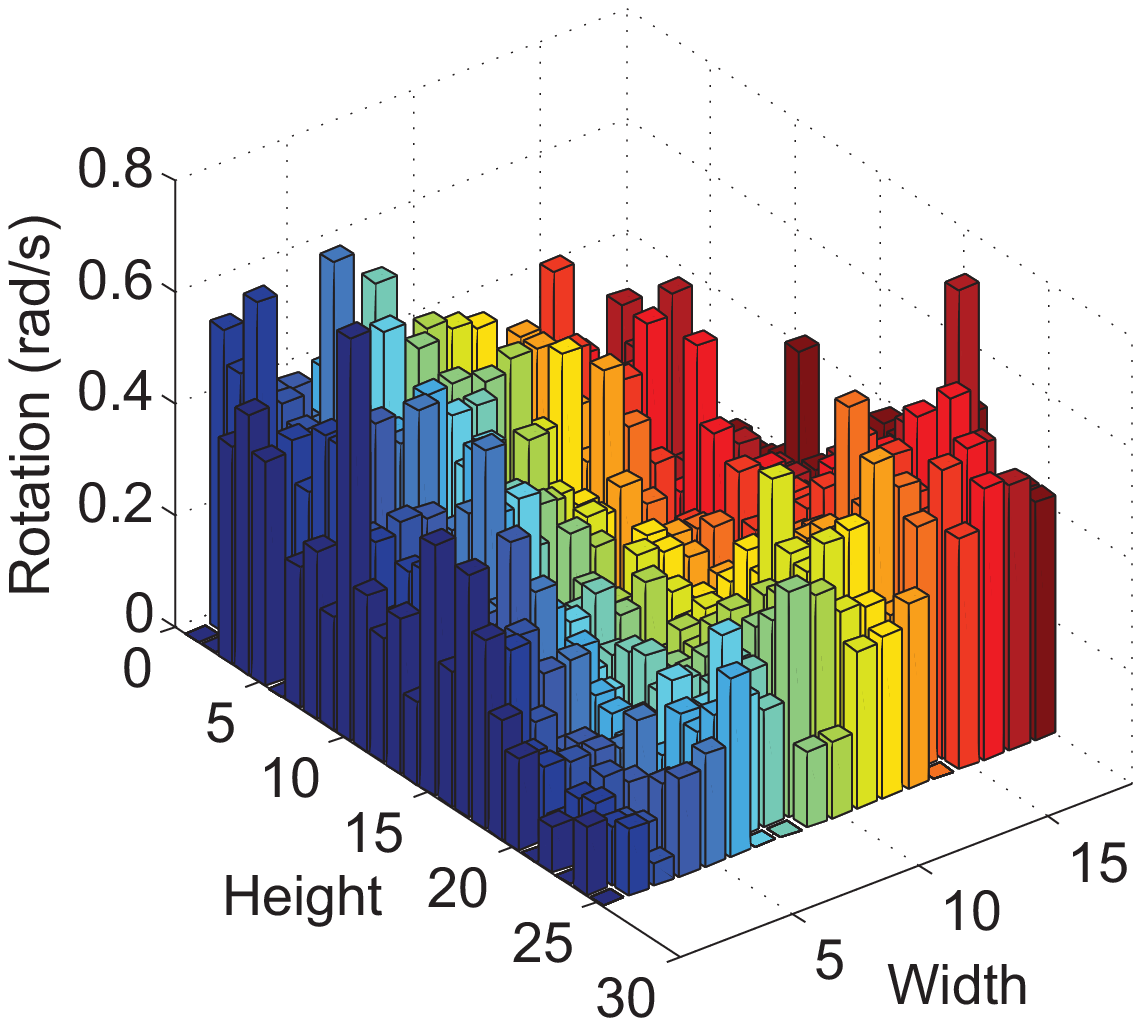}}
\vspace{-0.1in}
\caption{The distribution of both vibration and rotation on
  touchscreen under a given holding gesture.}
\label{fig:dis_acc_gyro}
\vspace{-0.1in}
\end{center}\vspace{-0.1in}
\end{figure}

We conduct a long period experiments to measure the vibration and rotation of the device
 with various touch coordinates.
We separate the touch screen into $25 \times 15$ small grids
 and calculate the mean vibration and mean rotation caused by tapping
 from each user for each grid-cell.
Figure~\ref{fig:dis_acc_gyro} shows the statistic device reactions from one of the users,
 who used to hold the lower part of the device by left hand,
 \ie the supporting point is near the left bottom,
 and taps the device by right index finger.
The experiments results show some interesting observations:
 (1) the amplitude of vibration and rotation depend on how the user holds the device.
    the father the coordinate from the holding point, the larger the vibration and rotation will be;
 (2) the changing trend of the vibration is obvious,
  leading to the possible holding position (which is also a behavior biometrics).

\begin{figure}[t]
\begin{center}
\subfigure[Accelerometer in Fling\label{fig:fling_acc}]{\includegraphics[scale = 0.25]{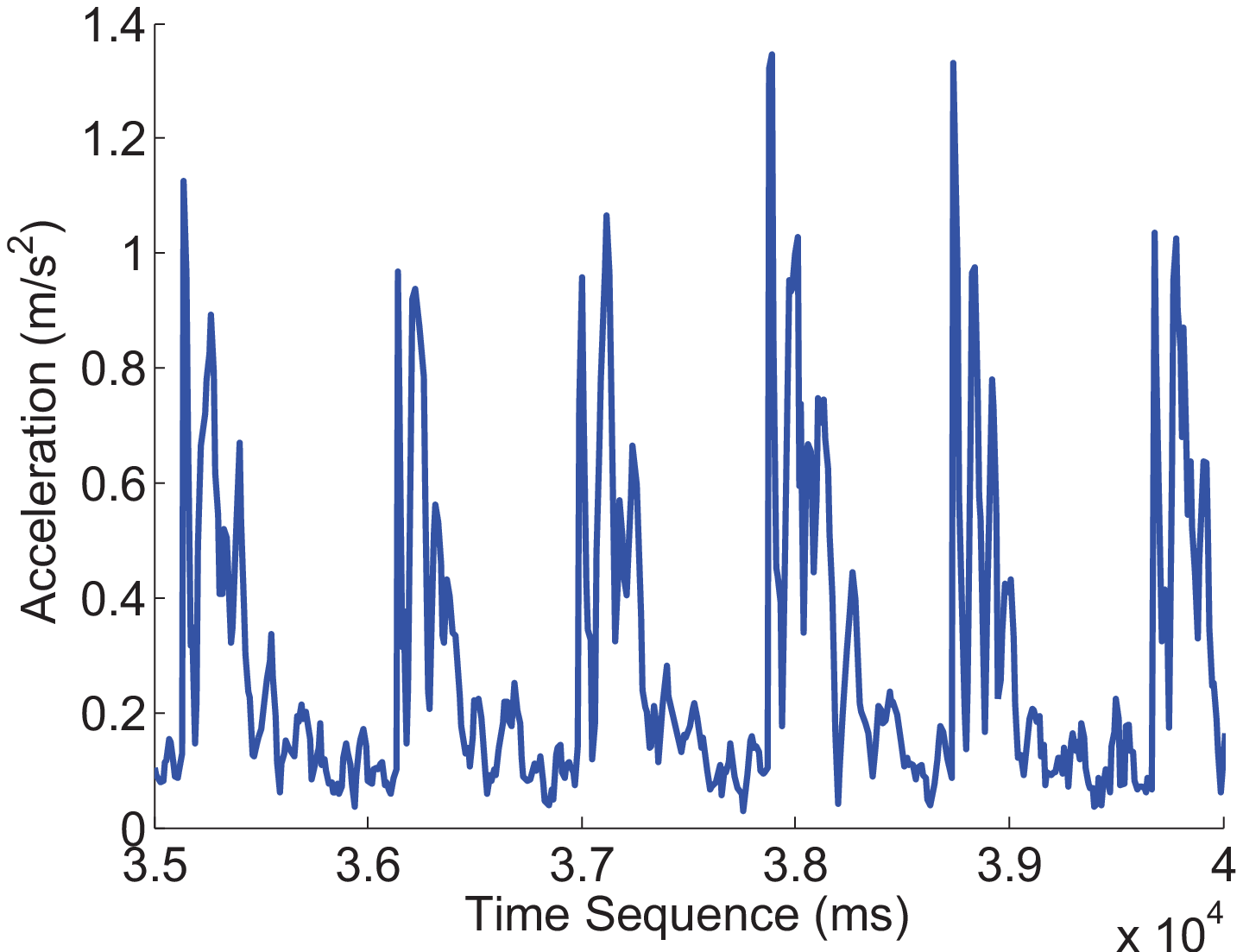}}
\subfigure[Gyroscope in Fling\label{fig:fling_gyro}]{\includegraphics[scale = 0.25]{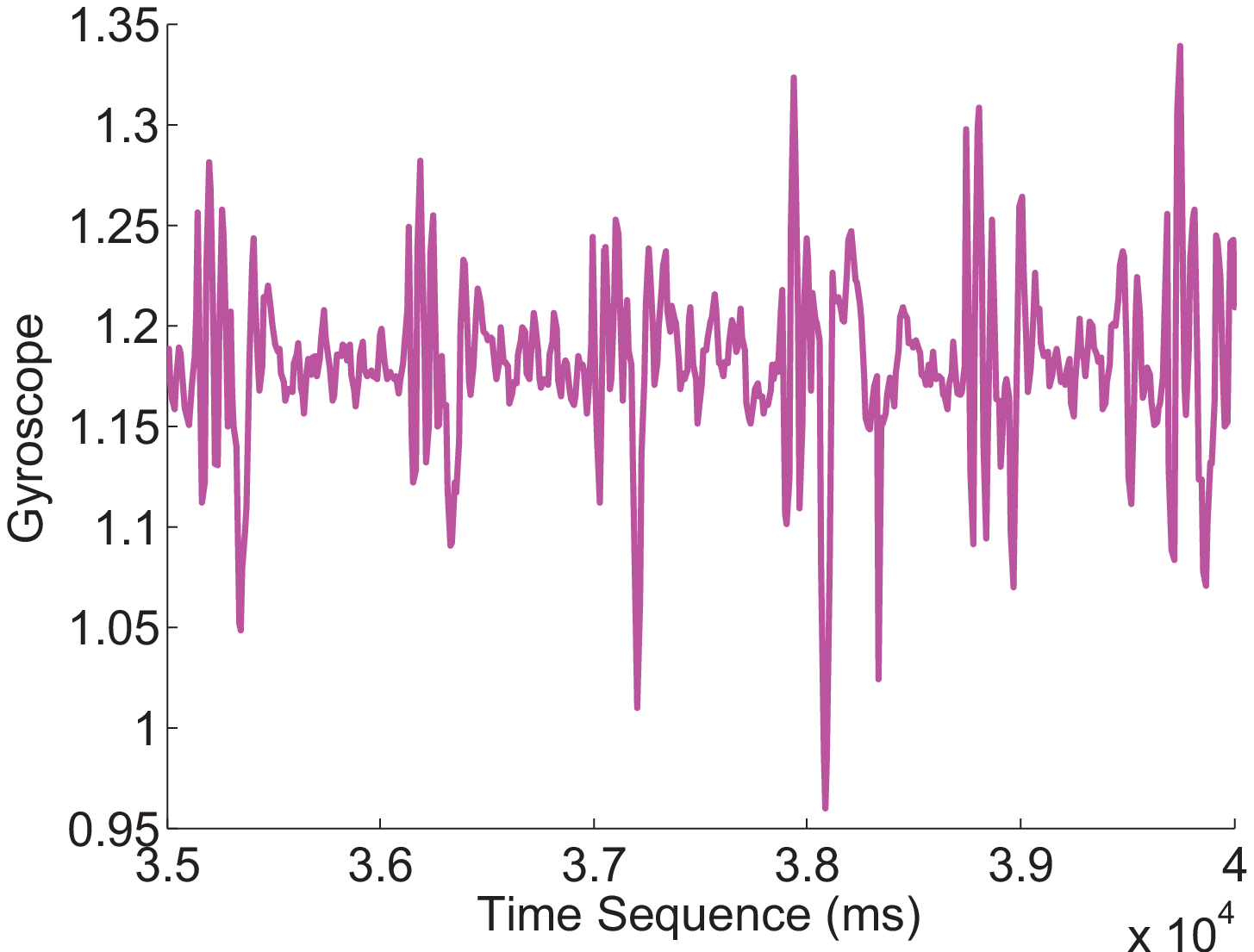}}
\vspace{-0.1in}
\caption{Sensor data when Fling}
\label{fig:fling_acc_gyro}
\vspace{-0.1in}
\end{center}
\vspace{-0.1in}
\end{figure}


We also investigate the device reaction to fling and scrolling
 and compare the reactions to the three gestures.
The results show that,
 reactions to both fling and scrolling are different from that to tapping,
 especially the amplitude.
Besides, both fling and scrolling do not drive the device to rotate
 in a large extent.


So far, we mainly consider the condition
 that the user is motionless or relatively still while the touch action.
In practice,
 a user may use the device while walking,
 the amplitude of the acceleration cause by walking
 is much larger than that from touch,
 which makes it infeasible to to extract the touch reaction feature in this case.
To address this challenge,
 we design a series of methods to extract
 the biometric walking feature for identify users in motion.

\subsection{Motion Analysis}

For the dynamic scenario,
 we analyze the motion features when the user uses the phone while walking,
 and combine the walking features with the interacting features (coordinate, duration
 and pressure) to construct the behavioral biometrics for identifying user in motion.

\begin{figure}[t]
\begin{center}
\subfigure[The FFT result of 4 step acceleration in the ECS. \label{fig:walking_fft}]{\includegraphics[scale =
    0.2]{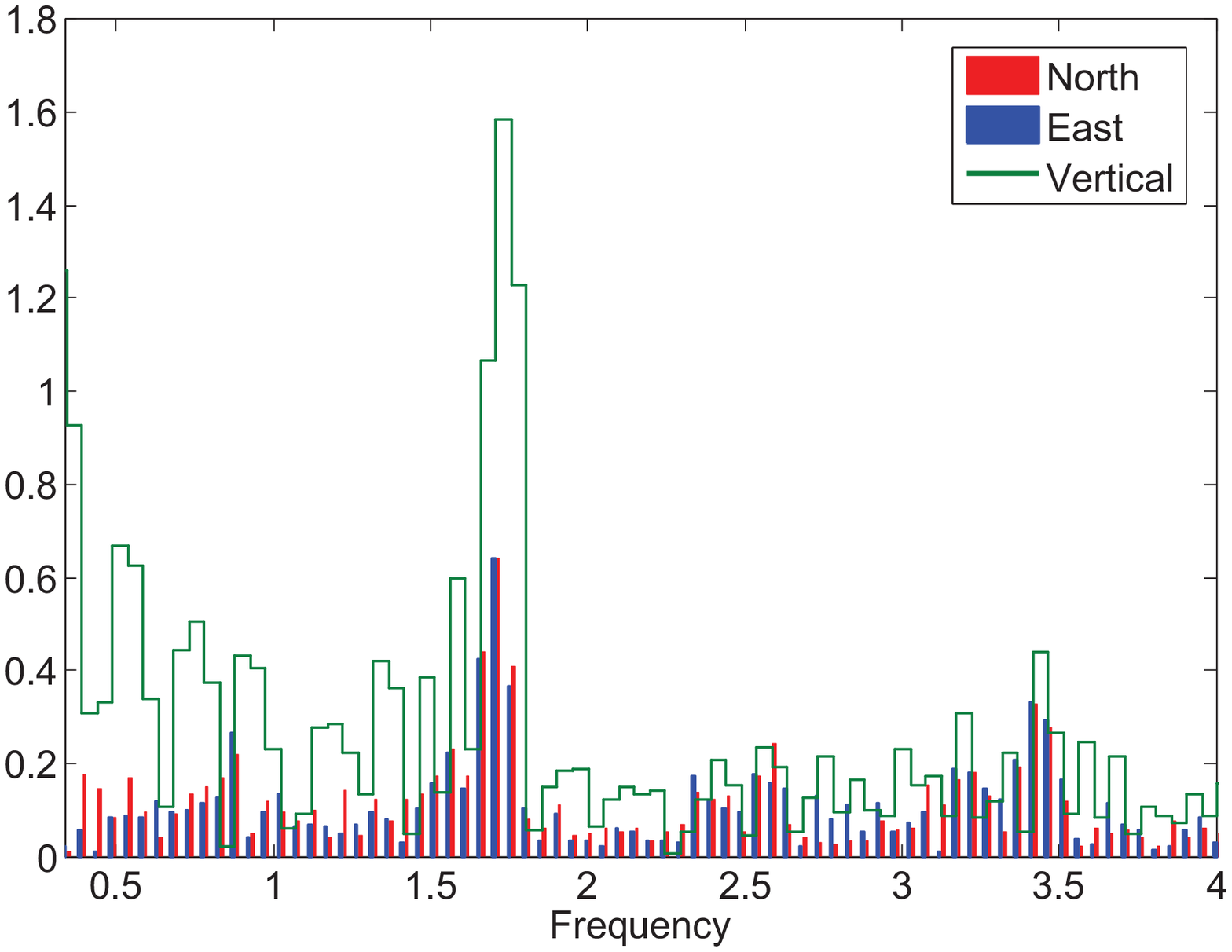}}
\quad
\subfigure[The raw vertical acceleration and filtered
  acceleration.\label{fig:walking_filter}]{\includegraphics[scale =
    0.2]{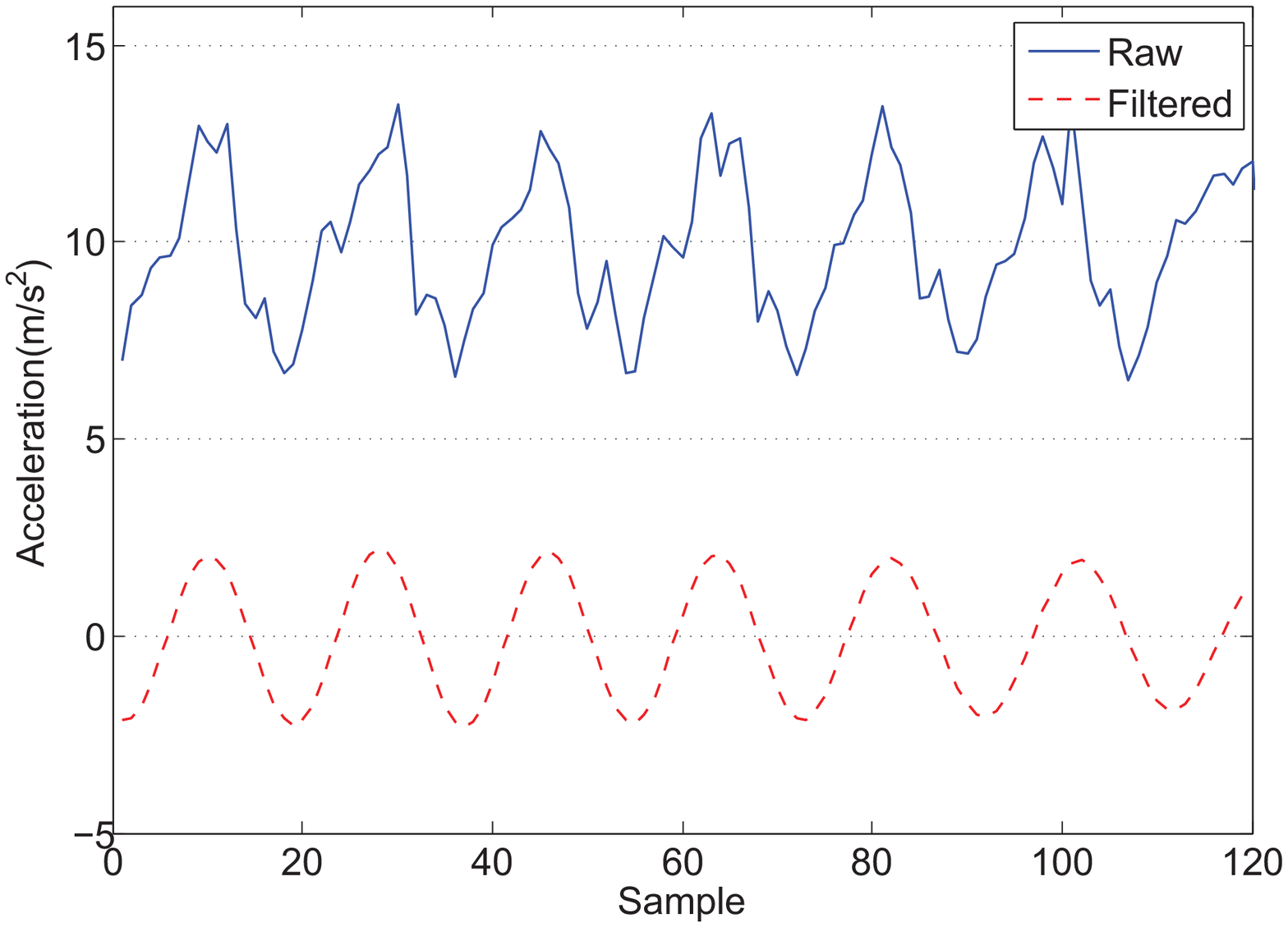}}
\caption{The frequency feature of acceleration in the earth coordinate
  system (ECS) while walking.}
\label{fig:walking_frequency}
\vspace{-0.1in}
\end{center}
\vspace{-0.1in}
\end{figure}

To accurately capture the walking features of different users,
 three steps are conducted in our method.
Firstly, considering a user could hold the phone in any attitude,
 we convert the raw acceleration vector in
 phone coordinate system (X,Y,Z axis)into the earth coordinate system (north, east, gravity)
 in real time.
Let the vector in the earth coordinate is $EA=\{EA_x, EA_y, EA_z\}$.
There are a lot of walking independent noise
 in the acceleration, which will greatly confuse
 the walking feature detection.
We analyze the acceleration while walking in the frequency domain,
 Figure~\ref{fig:walking_fft} shows that,
 the energy mainly locates around 2Hz,
 which is the user's walking frequency.
The energy in other frequency comes from noise.
To extract the pure walking acceleration,
 secondly, we filter $EA$ with a band pass filter
 to generate $EA'$.
Then we get the vertical acceleration $EA_v= EA'_z$ in the gravity
 orientation and horizontal acceleration $EA_h=\sqrt{EA_x^{'2} + EA_y^{'2}}$.
Figure~\ref{fig:walking_filter} shows the filtered vertical
 acceleration.
A simple step detection algorithm can be performed on the
 filtered vertical acceleration in real time.
Thirdly, we extract the walking feature from the processed
 acceleration data.
The vertical displacement of a walker is directly
 correlated to his/her stride length and height,
 hence it is an important feature.
Besides, 
 the step frequency and horizontal acceleration pattern also vary
 with different users.
To sum up, we extract four features of walking from $EA_v$ and $EA_h$:
 (1) Vertical displacement of each step by double integration of $EA_v$;
 (2) Current step frequency, calculated by the duration of each step;
 (3) Mean horizontal acceleration for each step;
 (4) Standard deviation of $EA_v$ for each step.

\section{Performance Evaluation}
\label{sec:evaluation}
We implemented \ourprotocol on Android phone as a service
 running background.
This service obtains the current app and touch events
 from system API, and captures sensory data from accelerometer
 and gyroscope.
We evaluate performance of \ourprotocol
 in different phases in both static and dynamic scenarios.
Android based HTC EVO $3$D and Samsung Galaxy S$3$ are employed in our experiments.
We have 100 volunteers, 10 of them are the smart phone owners,
 and other $90$ volunteers use their phones as guest users.
In average, for each owner, there are about 50 guests.
Several types of apps are considered sensitive in our experiments:
 message, mail, album, contacts and social networking apps.
We divide touch actions into three categories by gestures,
 including tap, scroll and fling.
In practice,
 the touch gestures are not limited to the three principle gestures,
 and there may be long press, double touch, pinch open, \etc.
But the fraction of such complex gesture is less than $5\%$ for daily usage.
As a result,
 we neglect other gestures in this experiments.
More than 50 actions of each gestures from each guest user
 are collected.
For each owner, thousands of actions of each gestures are collected.

\subsection{Identification in Static Scenario}
\begin{figure*}[!hptb]
\centering
\subfigure[Duration\label{fig:touch_dur_dif}]{\includegraphics[scale =
    0.25]{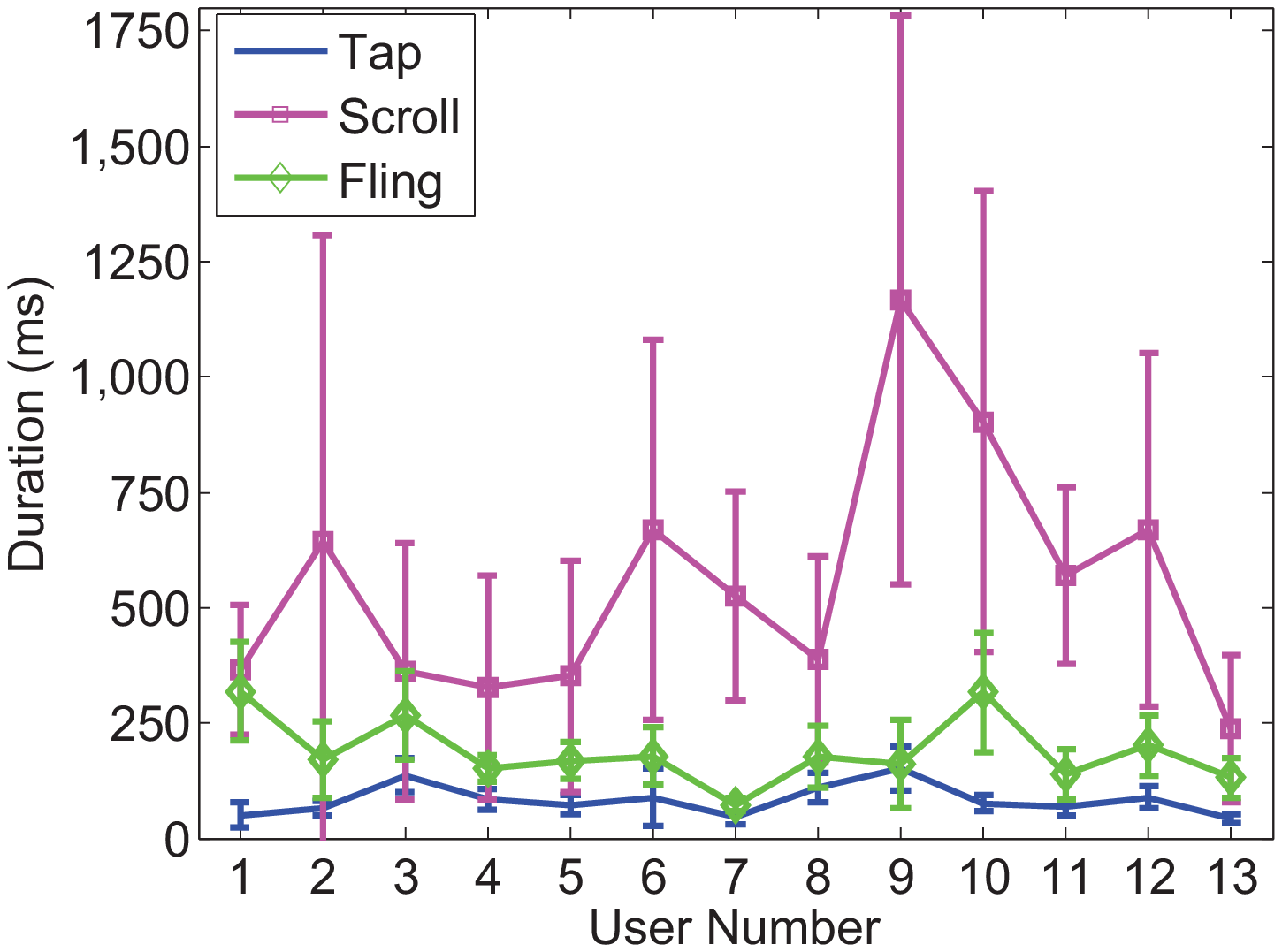}}
\quad
\subfigure[Pressure on
  Screen\label{fig:touch_pre_dif}]{\includegraphics[scale =
    0.25]{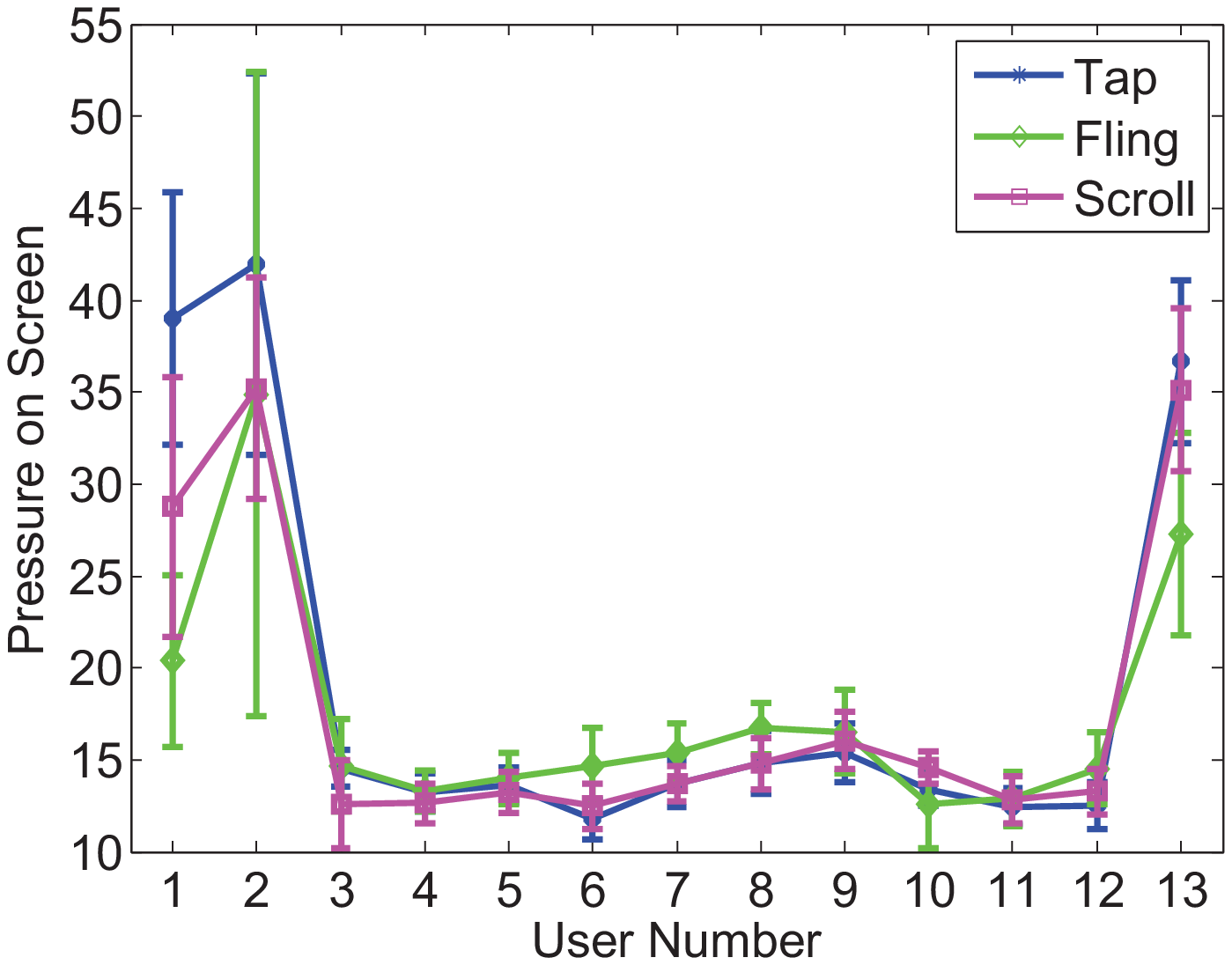}}
\quad
\subfigure[Device Vibration\label{fig:touch_vib_dif}]{\includegraphics[scale = 0.25]{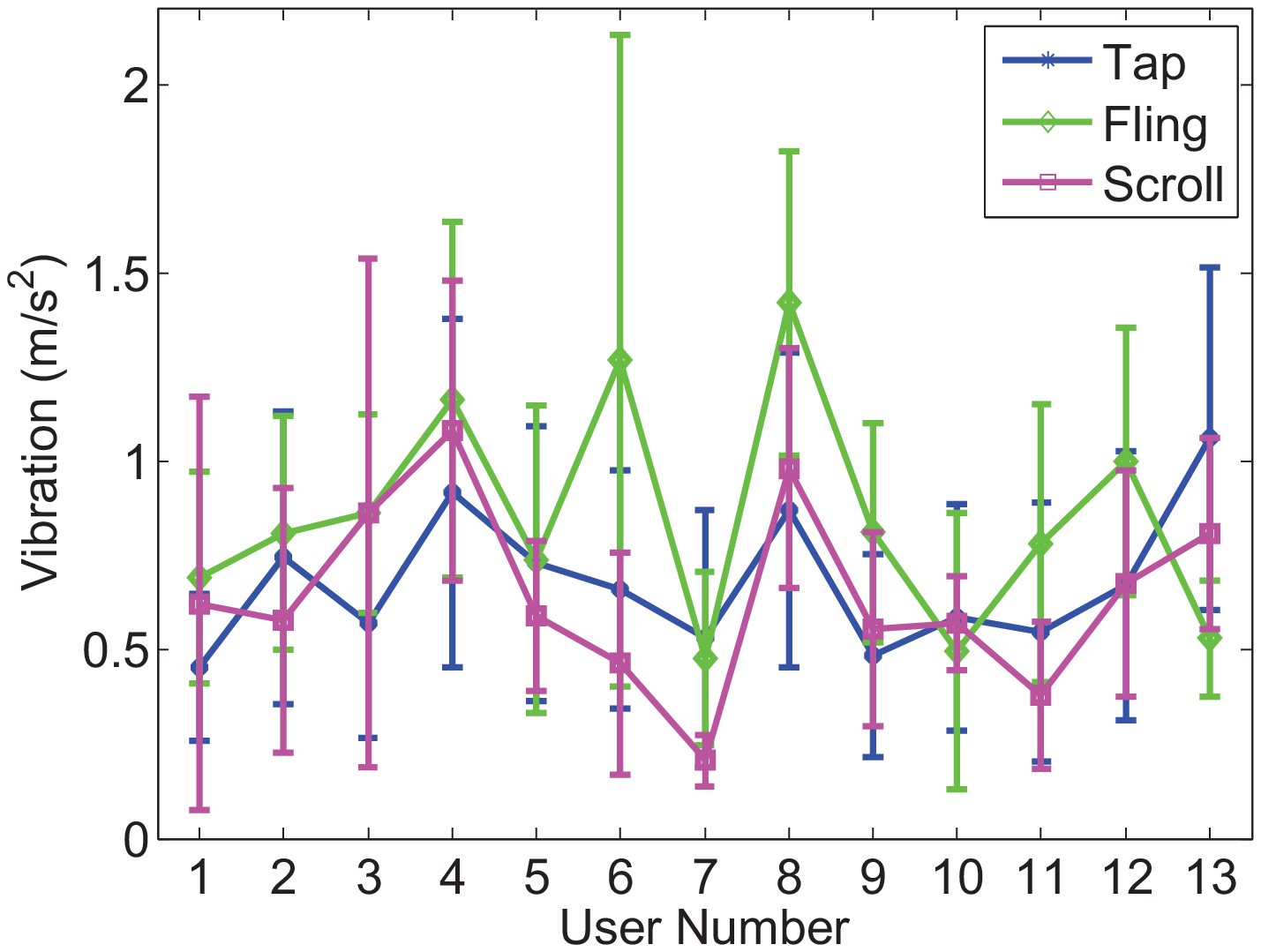}}
\quad
\subfigure[Device Rotation\label{fig:touch_rot_dif}]{\includegraphics[scale = 0.25]{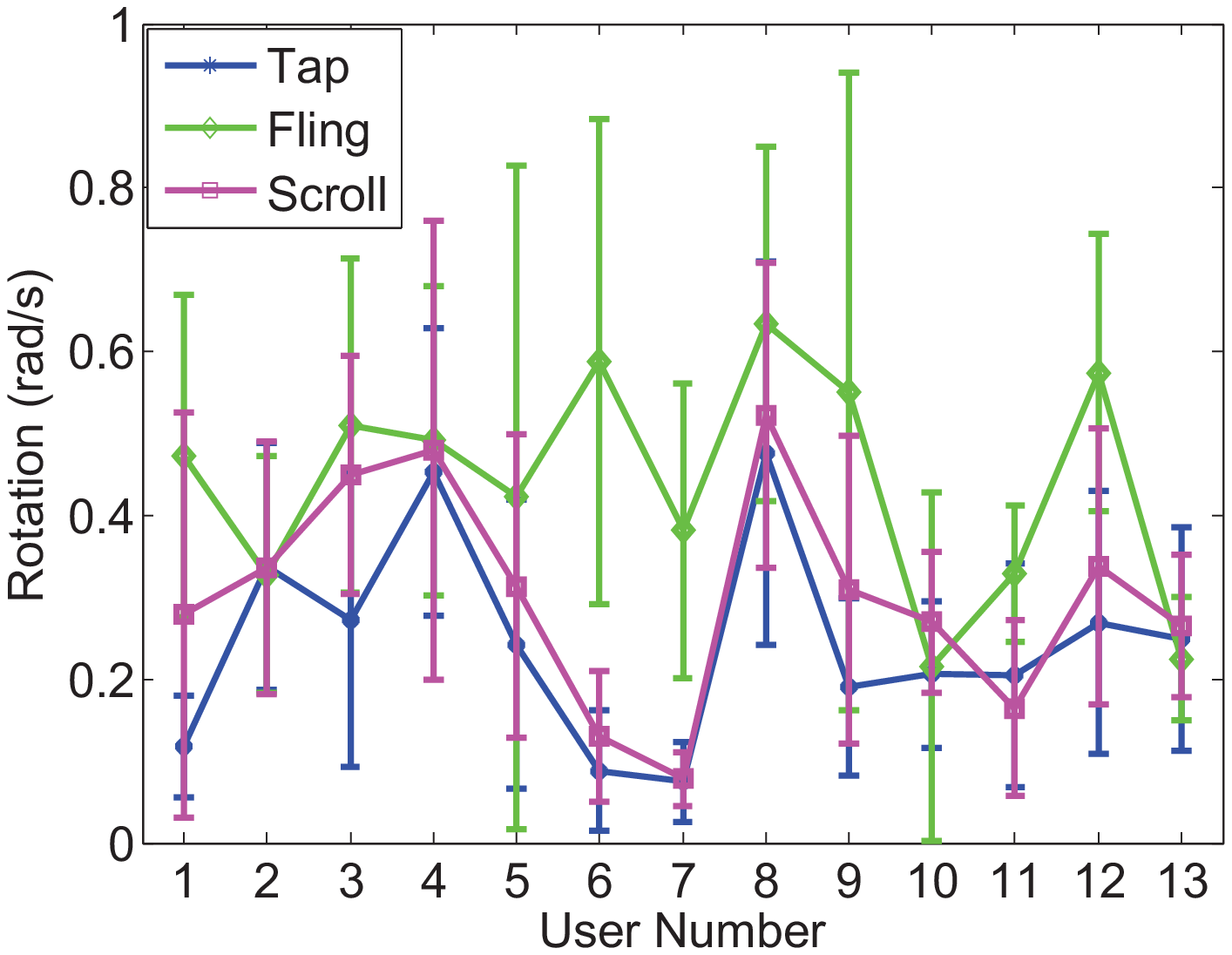}}
\caption{Touch event features for different users.}
\label{fig:touch_user_dif}
\end{figure*}
First, we explore the uniqueness of the behavior biometric in the static scenario.
With more than 100 actions for each user,
 we analyse the key features of users extracted from both the touching
 behavior and reaction of smartphone.
The analysis shows that there exit big diversity of each interacting feature
 among different users's.
The diversity mainly comes from the habits and biometric features of users,
 and the combination of multiple features provide
 unique user features.
For example, the mean duration of scrolls gesture of 100 users varies from 200 ms to 1200 ms,
 and the touch pressure varies from level 2 to level 40.
Figure~\ref{fig:touch_user_dif} shows 13 randomly selected users' interacting features
 of three types of gestures and present the diversity explicitly.



\begin{figure}[tb]
\centering
\subfigure[FAR\label{fig:in_app-far}]{\includegraphics[scale = 0.25]{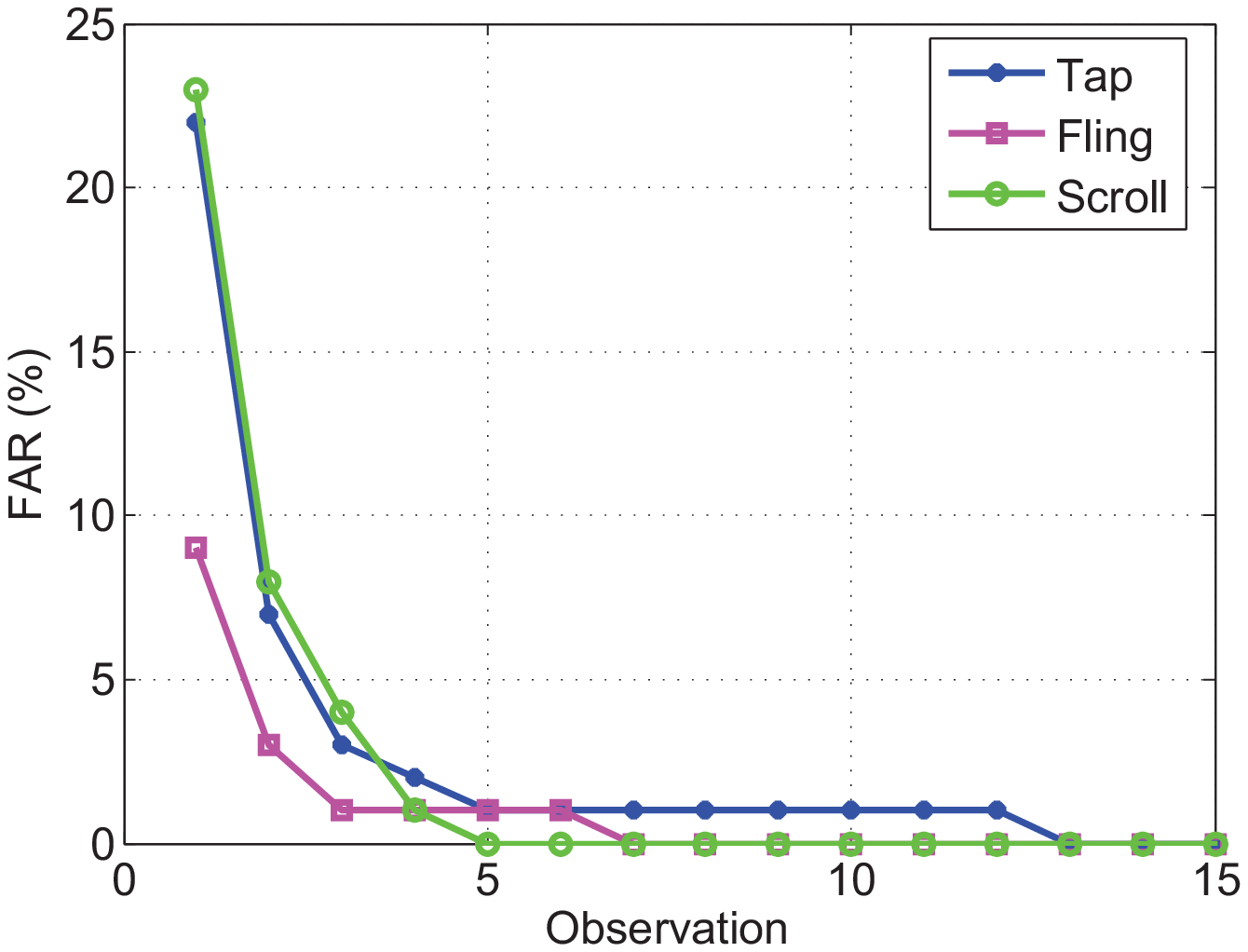}}
\subfigure[FRR\label{fig:in_app-frr}]{\includegraphics[scale =
    0.25]{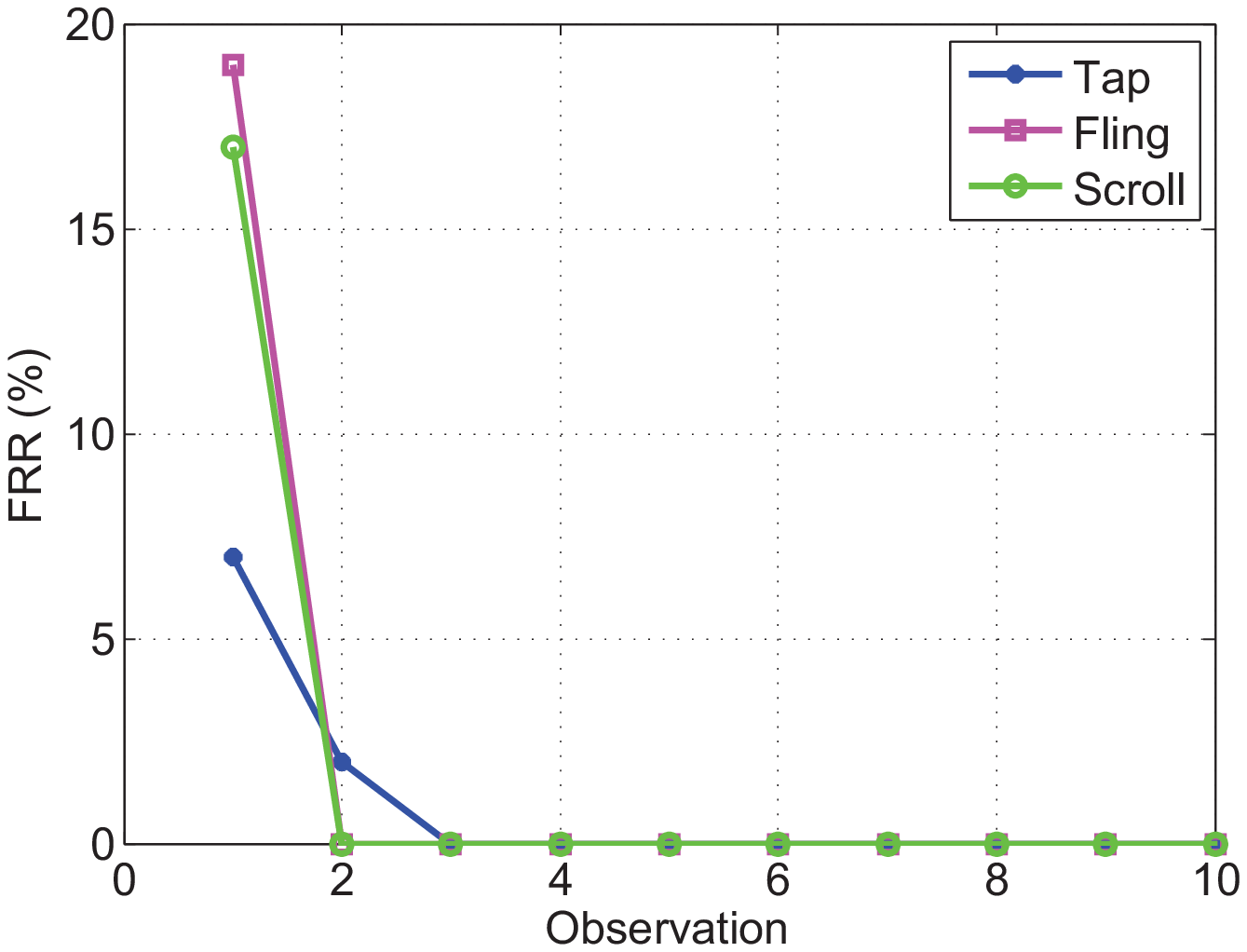}}
\caption{FAR, and FRR by
  different actions and different number of actions observed.}
\label{fig:inapp_result_all}
\end{figure}

Then we evaluate the performance of identification by three types of gestures,
 tap, scroll, and fling.
The main difference between scroll and fling is the moving speed
 and distance of the finger.
Fling is much faster while the distance of scroll is longer.
In \ourprotocol, observations and identification are made
 based on the three touch gestures.
Figure~\ref{fig:inapp_result_all} presents
 the false acceptance ratio (FAR), and false rejection ratio (FRR) of
 identification conclusion by different gestures with different number of
 observations.
Here the FAR is defined as the ratio of the number of identifications
 misjudge a guest as an owner over the total number of guest actions;
 and FRR is defined as the ratio of the number of identifications
 misjudge the owner as a guest over the total number of owner actions.
The results of 100 users show that,
 as illustrated in Figure~\ref{fig:in_app-far},
 the mean FAR of identification by one observation of tap is $22\%$, by one fling action is $9\%$,
 by one scroll action is $23\%$.
The FAR is reduced to below $1\%$ after observing about $3$
fling actions and with about $13$ observations
 the FAR achieves $0$ for all three gestures.
Surprisingly, Figure~\ref{fig:in_app-frr} shows that FRR  almost achieves $0$ with
 only $2$ observations for each gesture.
The experiments result show great discrimination
 of three gestures based on multiple features extracted by \ourprotocol.


\begin{figure}[t]
\centering
\subfigure[Guest Accuracy\label{fig:guest_accuracy}]{\includegraphics[scale = 0.25]{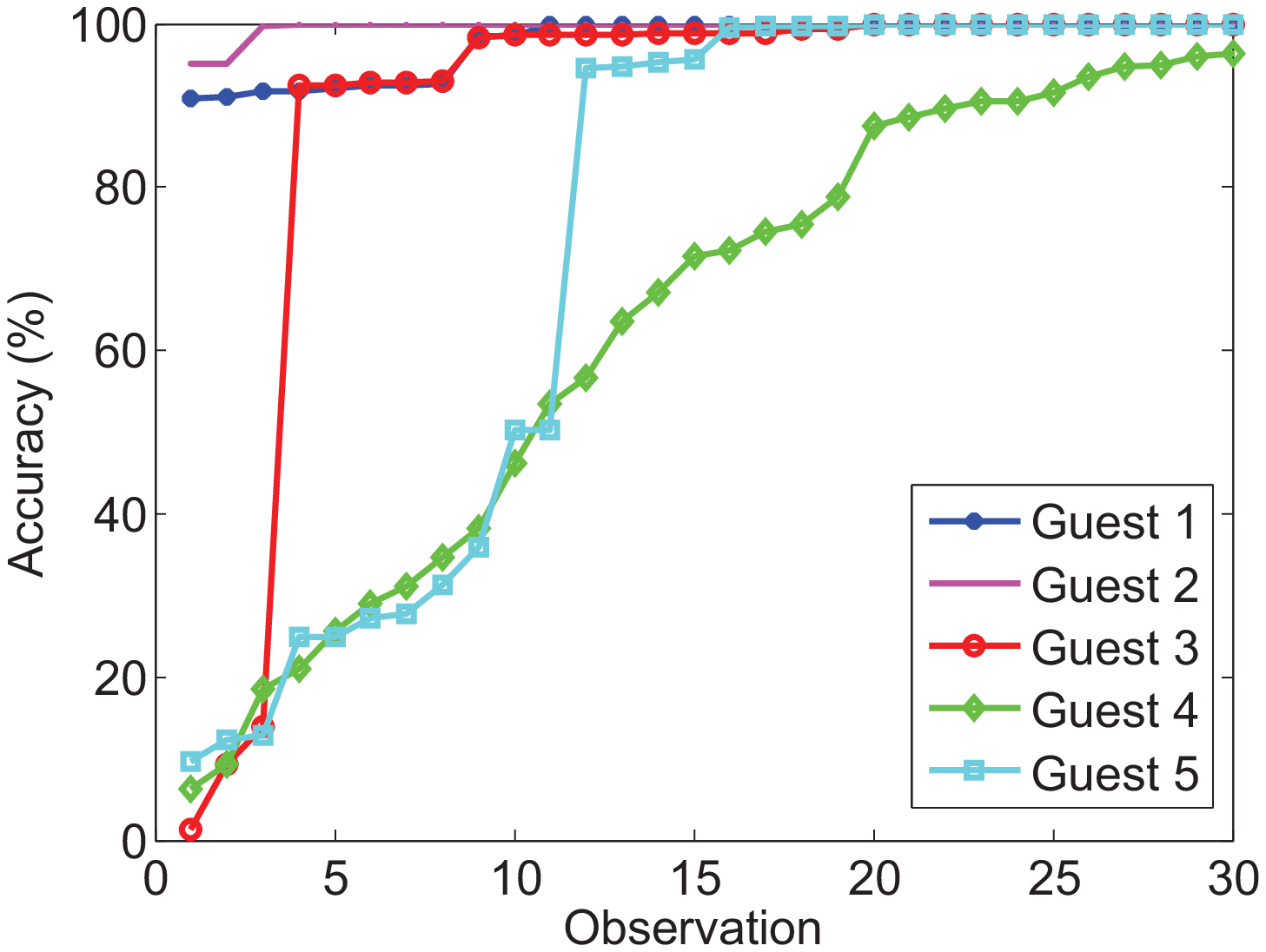}}
\subfigure[Owner Accuracy\label{fig:owner_accuracy}]{\includegraphics[scale = 0.25]{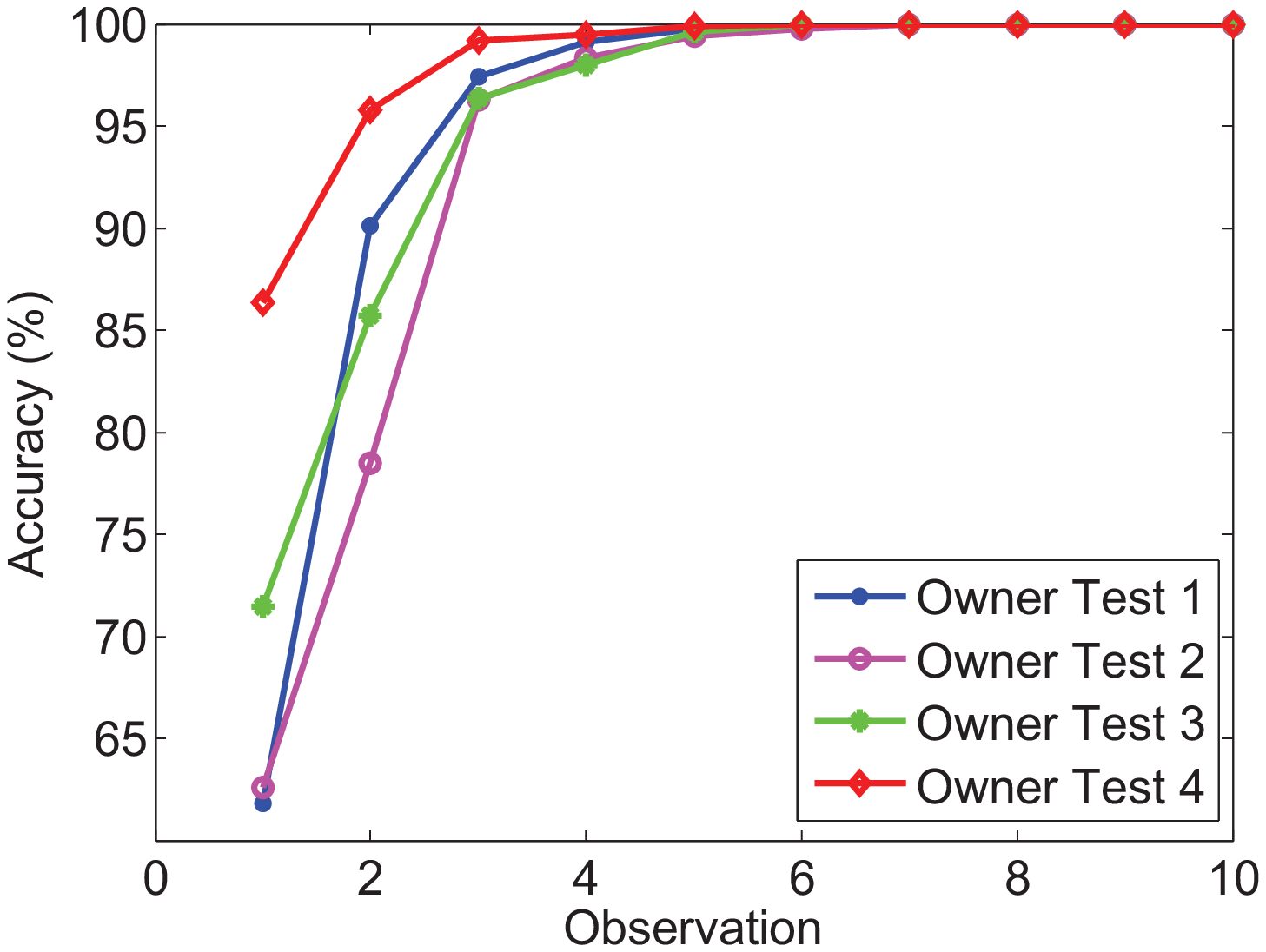}}
\caption{Identification based on sequence of random gestures.}
\label{fig:id_all}
\end{figure}
Now, we evaluate the performance of \ourprotocol\ in a more general scenario,
 with $100$ users interacting with smartphones freely as their daily usages,
 \ie, the action sequences are random combinations of three types of gestures.
Initially, the smartphone does not have any guests' behavior data,
 the framework could only identify
 the current user by \emph{one-class SVM} model.
Our experiments show that the initial accuracy is only $72.36\%$
 with one observation and the FAR is $24.99\%$.
However, 
 when a \emph{two-class SVM} model is
 trained with increasing amount of guest data,
  \ourprotocol reach high accuracy of identification with a small amount of observations.
Since the amount of training behavior data for the guest user
 is much smaller than that of the owner's data set,
 our experiments show that it is much faster to achieve a high accuracy
 for identifying the owner than identifying the guest.
Even so, the framework could reach over a $80\%$ accuracy within 10
 observations for identifying a guest.
Figure~\ref{fig:guest_accuracy} takes the results from five random selected guests
 and plots how soon the framework could identify the guest.
Similarly, as shown in Figure~\ref{fig:owner_accuracy},
 the owner will be identified with in $6$ observations.
Overall, in a general scenario,
 with only one observation, the FAR and FRR are about
 $20\%$.
But, with about $12$ observations of various actions, the FAR and
 FRR are both reduced to nearly $0$,
 meaning that there is no incorrect identification.


\subsection{Identification in Dynamic Scenario}

\begin{figure*}[!hptb]
\centering
\subfigure[Step Duration\label{fig:step_duration}]{\includegraphics[scale = 0.25]{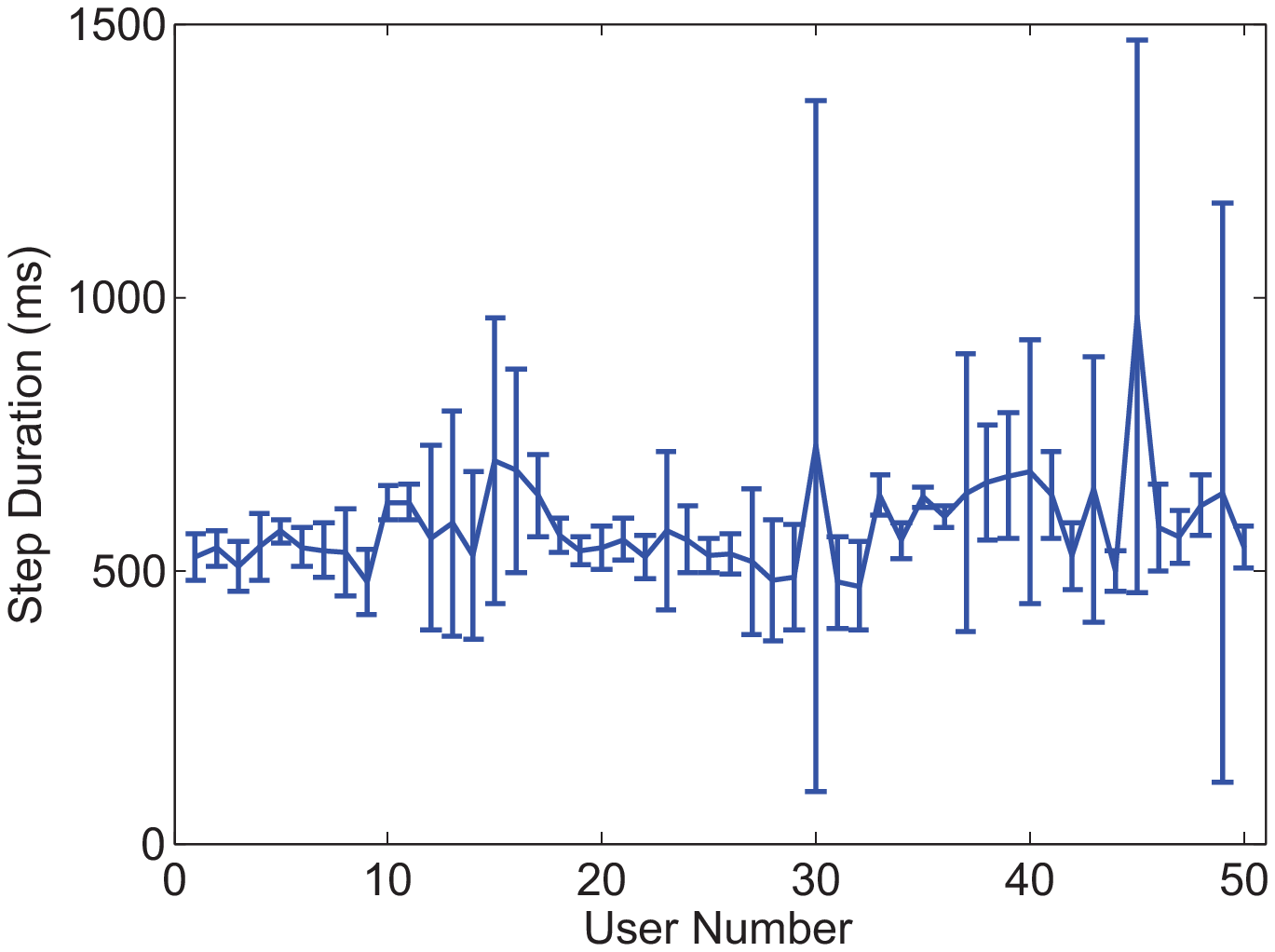}}
\subfigure[Vertical Displacement\label{fig:vertical_dis}]{\includegraphics[scale = 0.25]{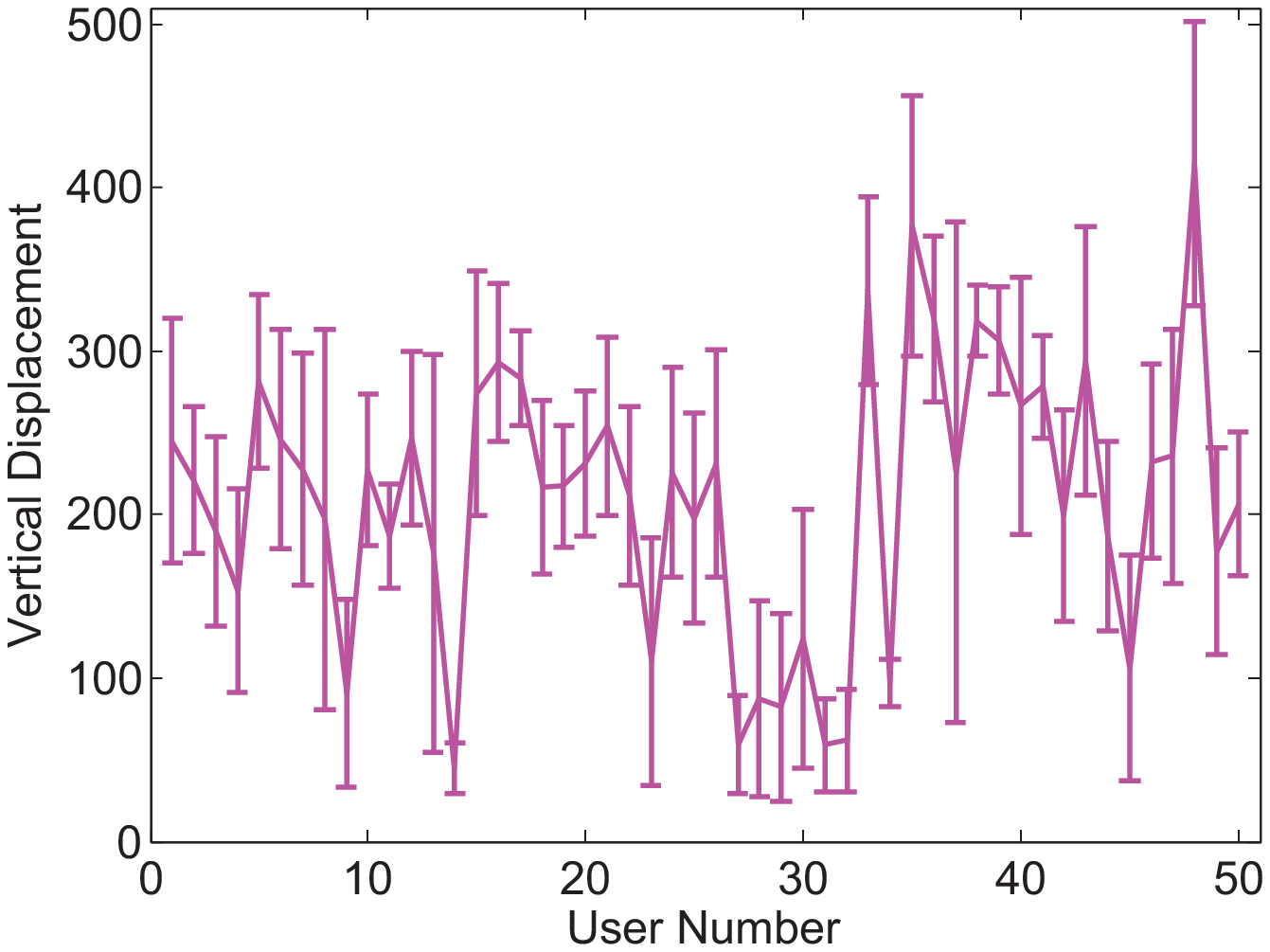}}
\subfigure[Horizontal Vibration\label{fig:hotizontal_vab}]{\includegraphics[scale = 0.25]{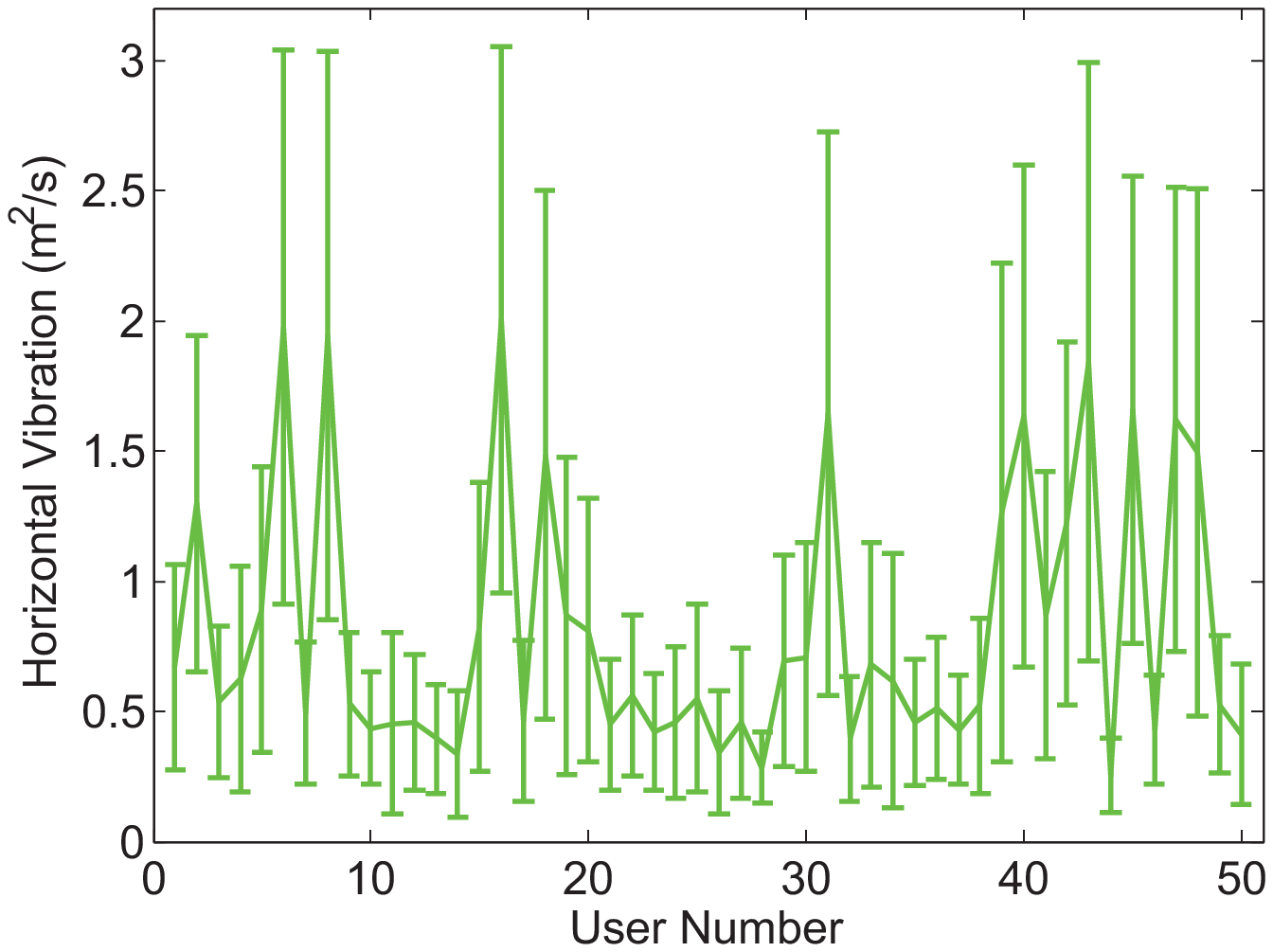}}
\caption{Walking features for different users.}
\label{fig:walking_user_dif}
\end{figure*}

\begin{figure}[t]
\begin{center}
\subfigure[Without walking features\label{fig:walking_no_feature}]{\includegraphics[scale = 0.25]{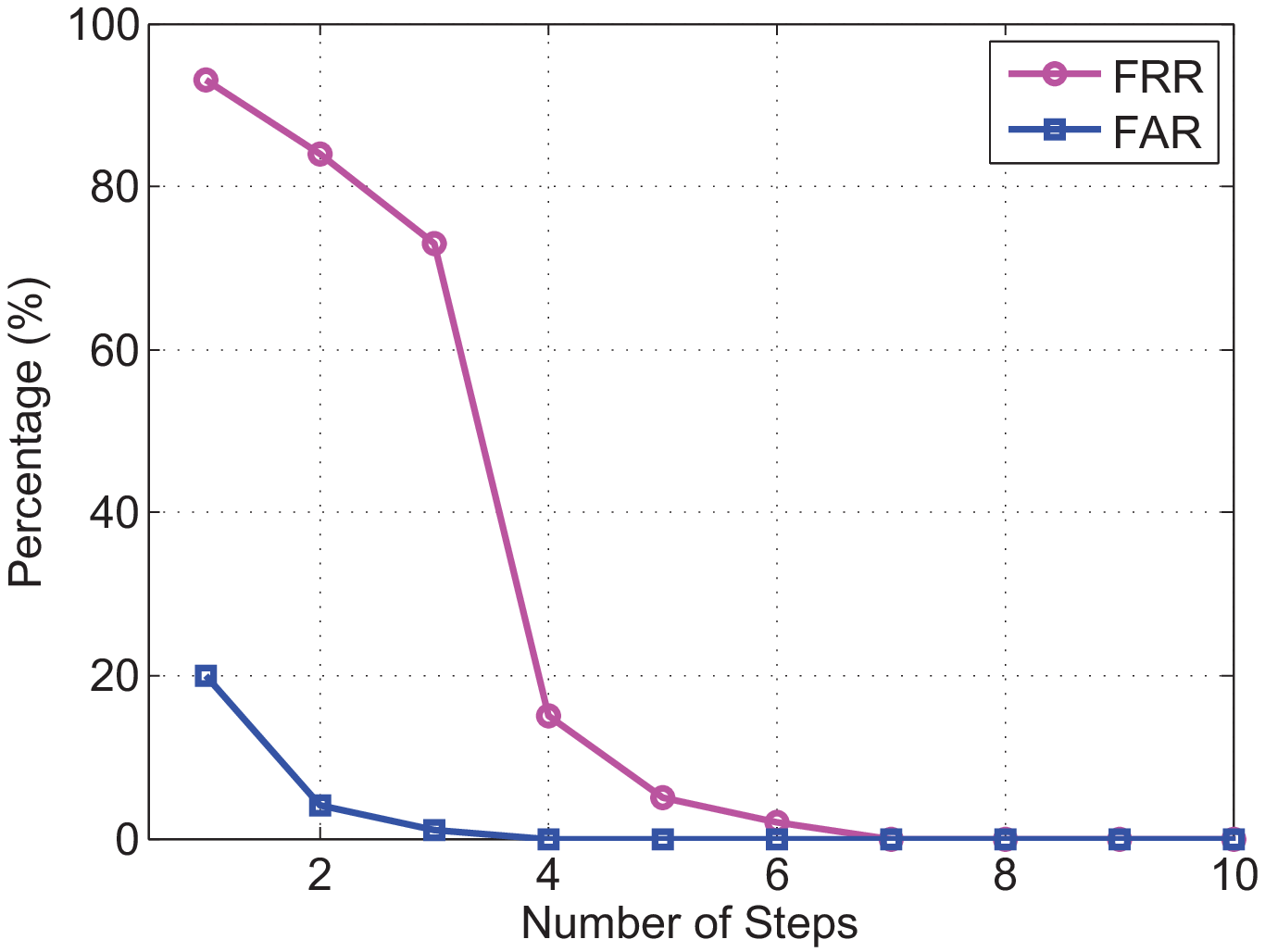}}
\subfigure[Use walking features\label{fig:walking_feature}]{\includegraphics[scale = 0.25]{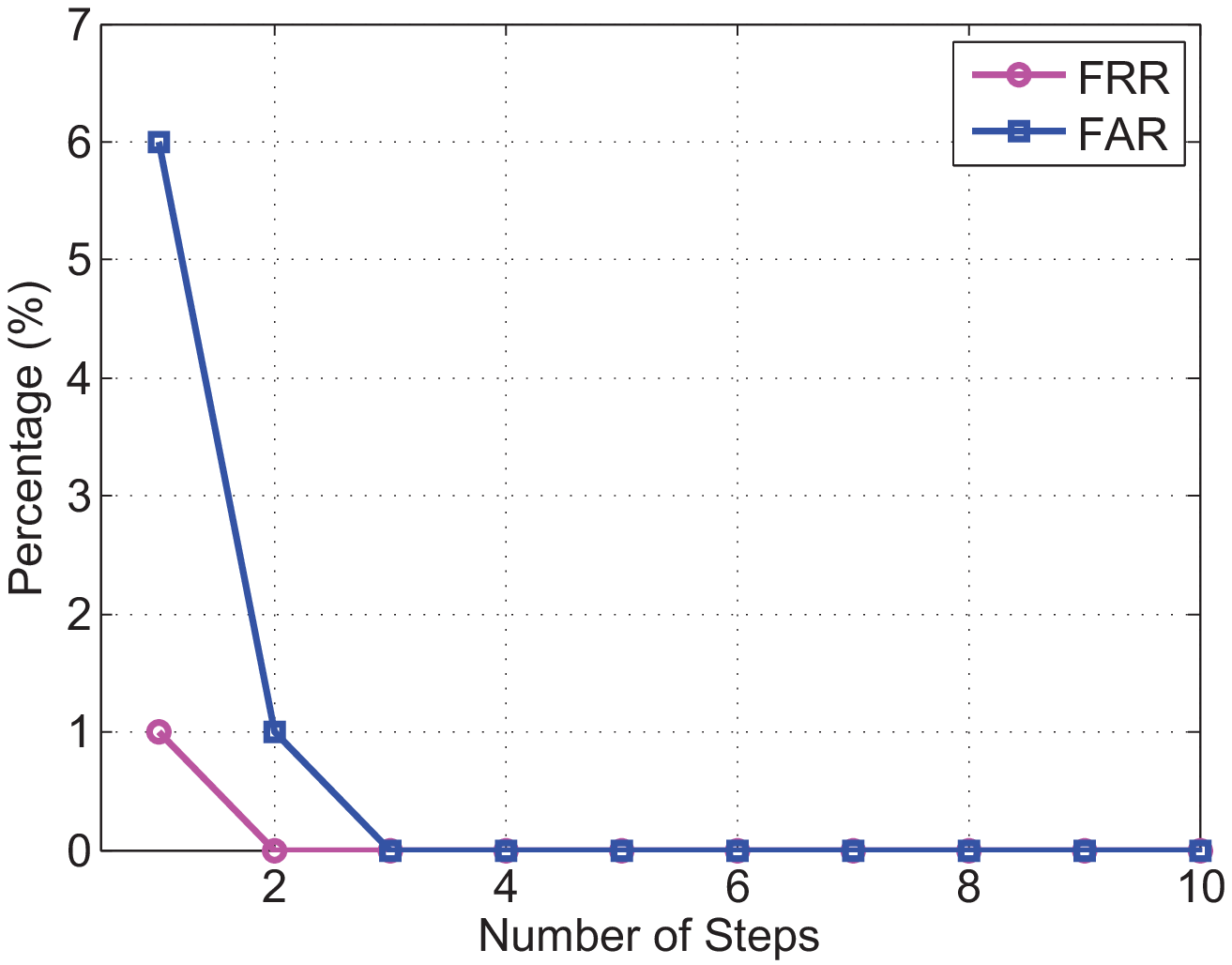}}
\caption{FAR and FRR by different number of steps observed.}
\label{fig:walking_far_frr}
\end{center}
\vspace{-0.1in}
\end{figure}

\begin{figure}[t]
\begin{center}
\subfigure[Guest Accuracy\label{fig:walk_guest_accuracy}]{\includegraphics[scale = 0.25]{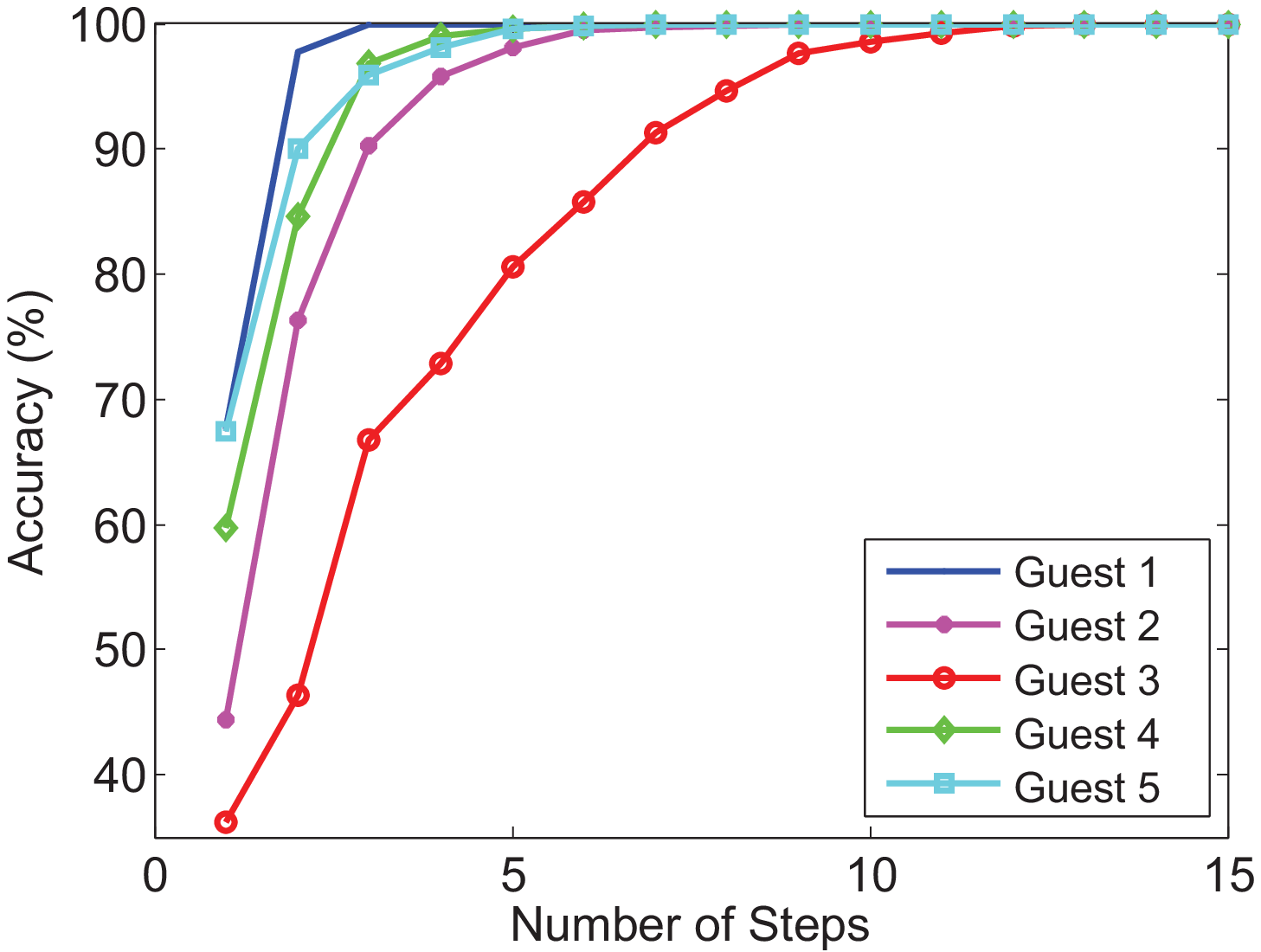}}
\subfigure[Owner Accuracy\label{fig:walk_owner_accuracy}]{\includegraphics[scale = 0.25]{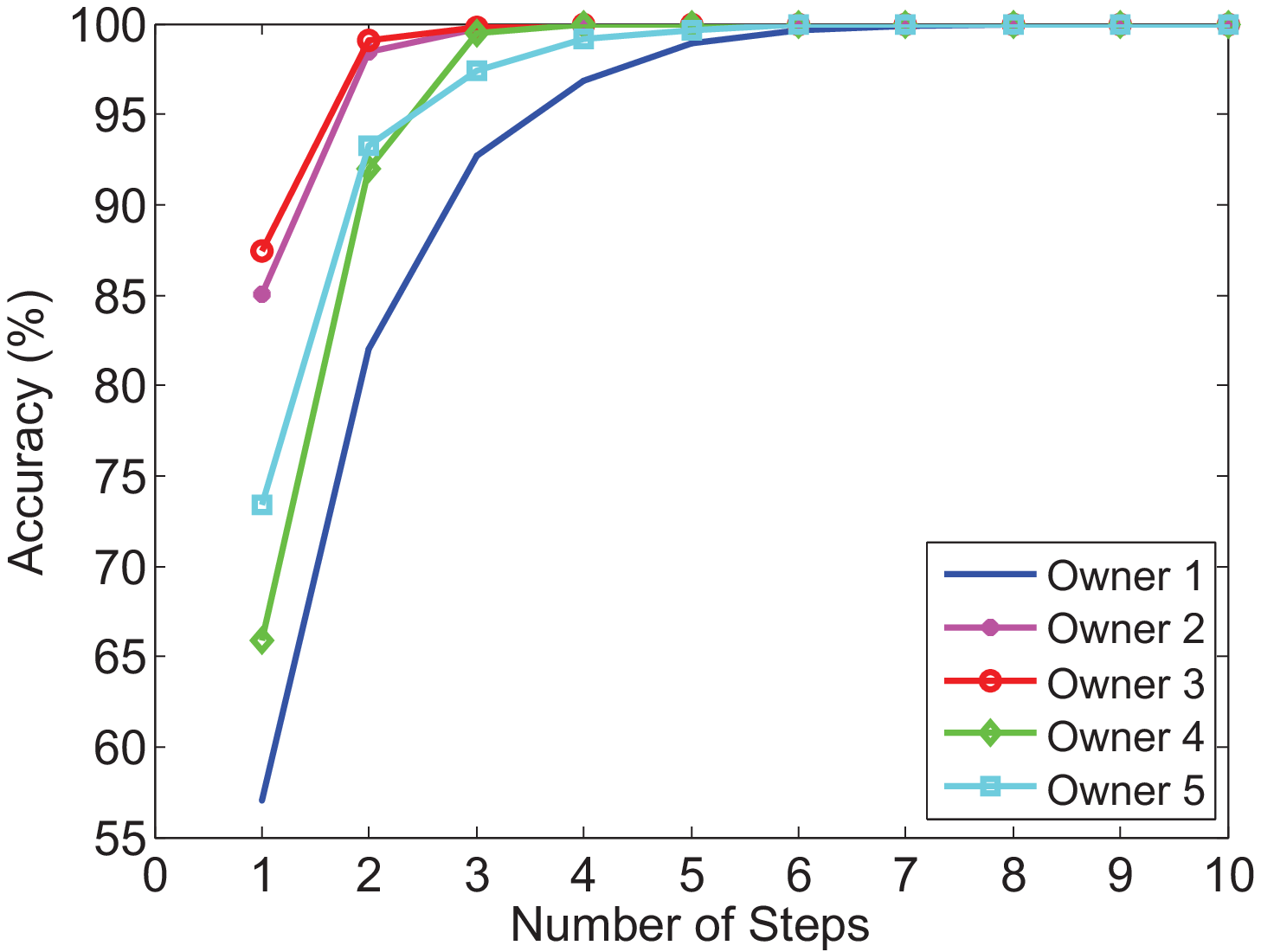}}
\caption{Identification based on walking feature.}
\label{fig:id_walk_all}
\end{center}
\vspace{-0.1in}
\end{figure}

In this part, users interact with smartphone while
 walking.
Recall that,
 the vibration and rotation reaction features caused by touch
 are no longer feasible due to the large movement cased by walking.
In this case, if only touch features (coordinate, pressure and duration)
 are used, as shown in Figure~\ref{fig:walking_no_feature},
 although the FAR reduces to $0\%$ after only 2 steps,
 the FRR is high as $18\%$ even after $4$ steps.
In the dynamic scenario,
 we extract 4 walking features,
 including vertical displacement, step duration, mean and standard deviation of the horizontal acceleration,
 from the filtered accelerations.
First, we explore the discriminative of walking features.
As shown in Figure~\ref{fig:walking_user_dif},
 the walking pattern varies greatly for different users,
 which give us an opportunity to identify the walking user rapidly.
So, we combined the walking features with touch features to establish the SVM model
 for dynamic scenario.
To evaluate the identification performance of \ourprotocol\ in dynamic scenario,
$50$ volunteers are required to use phones
 while they are waking freely.
We presents the FAR and FRR of identification results
 in Figure~\ref{fig:walking_feature}.
Our experiments show that the FAR and FRR reduce to $0$
 after only about $3$ steps.
Considering the different amount of training data,
 Figure~\ref{fig:id_walk_all} presents the achieved identification accuracy
 for randomly selected owners and guests separately.
After $12$ steps, the accuracy to identify a guest can achieve $100\%$,
 and after $7$ steps, the accuracy to identify the owner can achieve $100\%$.

\begin{figure}[!hptb]
\begin{center}
\subfigure[Adaptive Observation Frequency\label{fig:adaptive}]{\includegraphics[scale = 0.21]{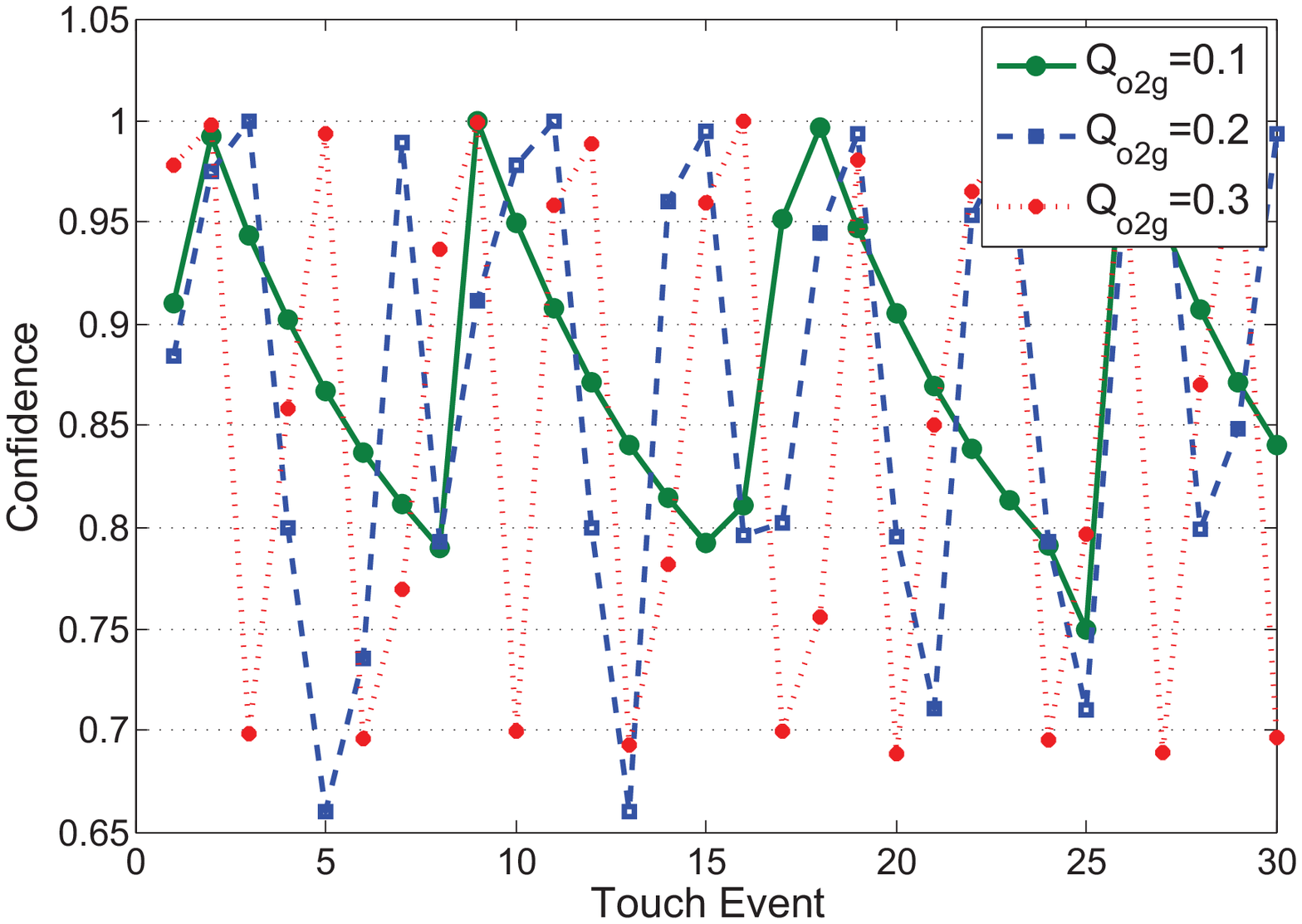}}
\quad
\subfigure[Energy Saved by Online Decision\label{fig:energysave}]{\includegraphics[scale = 0.21]{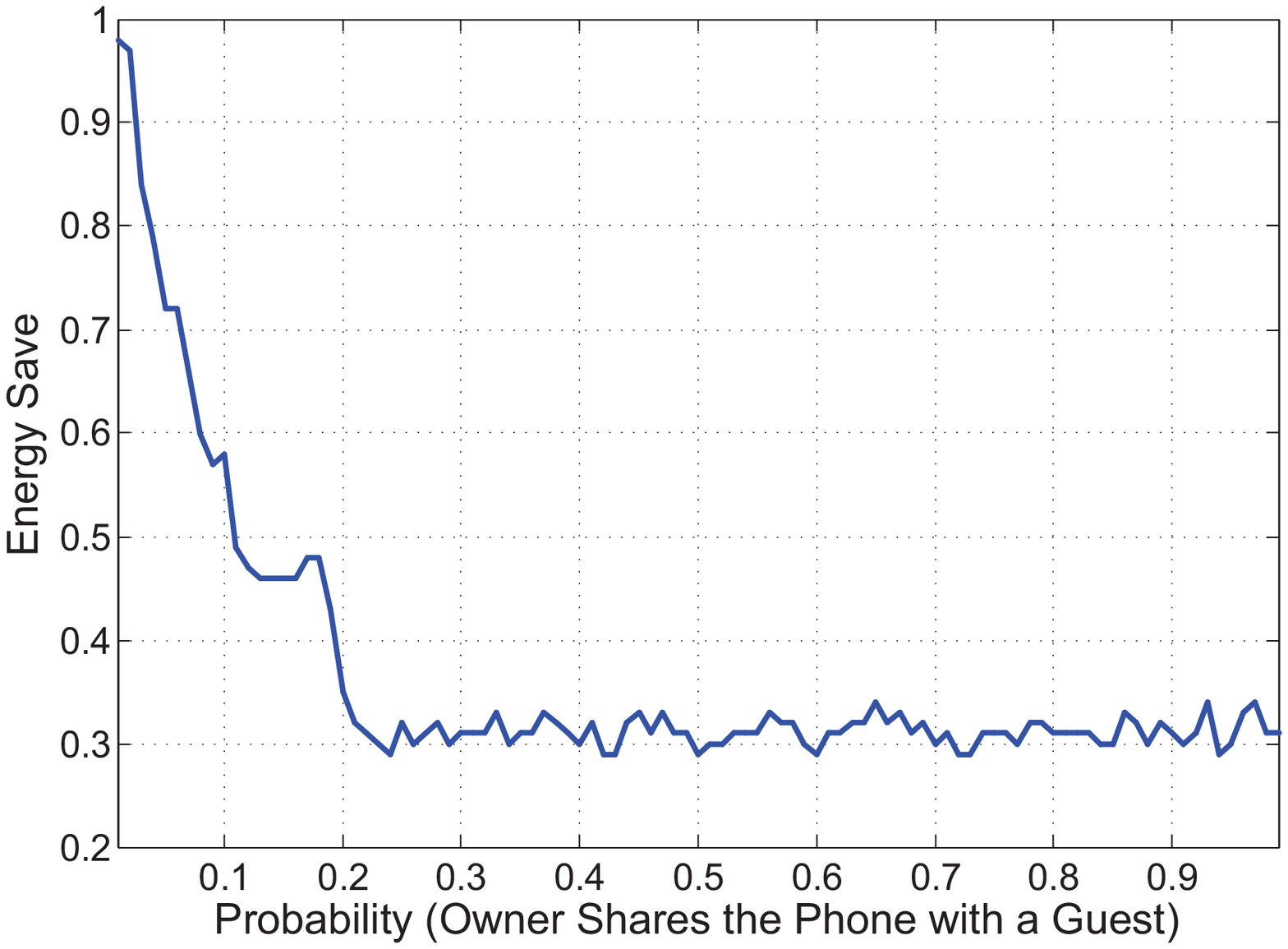}}
\caption{Online decision performance.}
\label{fig:onlinedecision}
\end{center}
\vspace{-0.15in}
\end{figure}

\subsection{Online Decision}
Using the historical observations,
 the probability $Q_{o2g}$ the owner shares the phone with a guest can be learned.
Based on the  our online decision strategy,
 the framework dynamically starts or stops observation,
 while keep a confidence about the current user's identity.
In the experiments,
 three users use phones with different phone share frequencies.
We set the confidence for a conclusion $\p_\theta=0.98$,
 and the acceptable estimation confidence $\p_\varphi=0.8$.
Figure~\ref{fig:adaptive} presents the confidence changing with observations.
During the rising edges, the observation has been started,
 while during the falling edges, the observation has been stopped.
The result shows that our online decision mechanism
 can successfully identify the user with averaged 2.26 actions delay
 with a $98\%$ accuracy guarantee when $57\%$ time the sensors are off
 for an owner who has $10\%$ probability to share the phone.
Figure~\ref{fig:adaptive} also presents adaptive observation frequencies
 yielded by our online decision strategy according to the owner's sociability.
While the observation stopped,
 motion sensors are off and no extra energy is cost by \ourprotocol.
Figure~\ref{fig:energysave} illustrate the energy saved by the online decision strategy
 for different $Q_{o2g}$.
When the owner rarely shares the phone,
 more than $90\%$ energy can be saved;
 while the owner share his/her phone frequently,
 the energy saving decreases and a relatively high observation frequency is necessary.

\section{Related Works}
\label{sec:related}

\vspace{-0.1in}
Most modern smartphones and tablets are equipped with multiple sensors,
 which have risks to expose the user's activities and privacy
 to attackers~\cite{egele2011pios,cai2009defending}.
For example, TapLogger~\cite{xu2012taplogger} deduces a user's tap input to a smartphone
 from its integrated inertial sensory data.
Other similar works include TouchLogger~\cite{cai2011touchlogger} and ACCessory~\cite{owusu2012accessory},
 which also utilize accelerometer readings to infer keystrokes.
As an improved work, TapPrints~\cite{miluzzo2012tapprints}
 infers the tap information on a smartphone or tablet from the combination
 of accelerometer and gyroscope for better accuracy.
And such framework also demonstrates the possibility of compromising user's privacy.

Researchers also inferred keystroke on traditional keyboard based on acoustic signal~\cite{asonov2004keyboard, zhuang2005keyboard}, timing event observation~\cite{foo2010timing},
 and electromagnetic waves~\cite{vuagnoux2009compromising}.
(sp)iPhone~\cite{marquardt2011sp} takes the advantage of motion sensors
 to detect the vibration and infer the keystroke of nearby keyboard.
Both~\cite{aviv2010smudge} and~\cite{zhang2012fingerprint} study the possibility of
 identifying the password sequence by examining the smudge left on the touch screen.
Besides, \cite{maggi2011poster} and \cite{raguram2011ispy} propose the method to
 infer a user's input by observing the touch action  with a camera.
These works indicate that the interaction between user and device
 can be observed through sensors and may cause a privacy leakage.


There are two categories of biometrics for identification users:
 physiological (such as fingerprints, facial features) and behavioral biometrics
 (such as speaking, typing, walking).
Physiological biometrics usually requires special recognition devices.
Some physiological biometrics, like face and voice can
 be detected by smartphones,
 but usually cost expensive computation and energy cost and have a
 high error rate.
For example, in \cite{koreman2006multi}, the (equal error rates) EER for
 face recognition is around $28\%$ and for voice is around $5\%$.
Keystroke is a popular behavioral biometric.
\cite{joyce1990identity} presented a survey on the large body of
 literature on authentication with keystroke dynamics.
Researchers also propose authentication token based mechanisms
 to identify legal users,
 \eg, wireless token~\cite{nicholson2006mobile}. 
However, they require additional hardware
 and are not convenient for daily smartphone usage.
On the smart phones with touch screens,
 PINs, pass-phrases, and secret drawn gestures are the
 commonly used authentication methods~\cite{dunphy2010closer}.
Such approaches are easily deployed with off-the-shelf smartphones,
 but they are vulnerable to shoulder surfing, smudge and other attacks.
Recently, there is a growing body of work that
 use the features of touch behavior to verify users.

\emph{Touch Behavioral Biometrics:}
Several existing approaches have used the touch behavior biometrics
 for various security purposes.
\cite{de2012touch} proposes an password application,
 by which the user draw a stroke on the touch screen as a input password.
 Pressure, coordinates, size, speed and time of the stoke are used to identify valid user.
 Overall accuracy of this work is $77\%$ with a $19\%$ FRR and $21\%$ FAR.
\cite{WM-CS-2012-06} uses four features (acceleration, pressure, size, and time)
to distinguish the true owner and impostor to enhance the security of passcode.
Its identification system achieves $3.65\%$ EER.
A user enters a  password by tapping several times on a touch surface with one or more fingers. PassChord failed to authenticate for $16.3\%$ of the time.
There are some other work addressing
 the user identification issue with touch features, \eg, \cite{seo2012novel}.
As we see, with pure touch data, there may be a high error rate.
Furthermore, the above work verify users in an explicit way,
 which works similar as inputting a pincode and are inconvenient for
 device share or multi-user scenarios.

Recently, there are some work address the
 user identification with behavior biometrics in
 continuous or implicit manner.
In those work, identification services run in background
 and identify the current user in real time.
For example, \cite{frank2012touchalytics} continuously authenticates users based on 30 behavioral features,
 including touch features and motion sensor features.
In this work, the EER is approximately $13\%$ with a single stroke
 and converges to a range between $2\%$ and $3\%$ with 11 to 12 strokes.
SenGuard~\cite{shi2011senguard} combines motion, voice, location history
 and multi-touch data to identify users of smartphone,
 whose average error rate is $3.6\%$.
FAST~\cite{feng2012continuous} uses a special digital sensor
 glove to achieve highly accurate continuous identification.
Those work use special devices or motion sensors to enrich the identification features
 to improve the poor accuracy with pure touch information.
But they ignore the scenario that the user uses mobile phone while walking,
 where the micro features caused by touch is suppressed
 by the large scale movement which will make
 the accuracy of the exiting methods deteriorate.

Here we propose \ourprotocol
 to identify users with a background service
 running continuously and implicitly in the off-the-shelf smartphone.
\ourprotocol achieves an high accuracy in both static and dynamic scenarios,
 which are not addressed in previous work.

\vspace{-0.08in}
\section{Conclusion}
\label{sec:conclusion}
In this paper, we present \ourprotocol, a framework to verify whether
  the current user is legitimate owner of the smartphone based on the
  behavioral biometrics, including touch behaviors and walking patterns.
We establish a model and a novel method to silently verify the
 user with high confidence:
 the false acceptance rate (FAR) and false rejection rate (FRR) could
 be as low as  $< 1\%$ after only collecting about $10$
 actions.
We have found that a user's touch signatures if used in
 conjunction with the walking patterns will achieve significant low
 error rates for user identification in a completely non-intrusive
 and privacy preserving fashion.

\vspace{-0.08in}
\small{

}

\end{document}